\documentclass[aps,pre,twocolumn]{revtex4-1}

\usepackage{amssymb,amsfonts,amsmath}
\usepackage{graphicx}
\usepackage{dcolumn}
\usepackage{bm}
\usepackage{mathrsfs}
\usepackage{mhchem}
\usepackage{subfigure}
\usepackage{color}

\usepackage{booktabs}
\usepackage{epstopdf}
\usepackage{multirow}

\usepackage{chemarrow}
\usepackage{extarrows}
\usepackage{mathtools}

\usepackage{epsf}
\usepackage{epstopdf}
\newcommand{\be}{\begin{equation}}
\newcommand{\ee}{\end{equation}}
\newcommand{\bea}{\begin{eqnarray}}
\newcommand{\eea}{\end{eqnarray}}

\begin{document}

\title{Active matter, microreversibility, and thermodynamics}

\author{Pierre Gaspard}
\email{gaspard@ulb.ac.be}
\affiliation{ Center for Nonlinear Phenomena and Complex Systems, Universit{\'e} Libre de Bruxelles (U.L.B.), Code Postal 231, Campus Plaine, B-1050 Brussels, Belgium \\ {\tt https://orcid.org/0000-0003-3804-2110}}

\author{Raymond Kapral}
\email{rkapral@utoronto.ca}
\affiliation{ Chemical Physics Theory Group, Department of Chemistry, University of Toronto, Toronto, Ontario M5S 3H6, Canada \\ {\tt https://orcid.org/0000-0002-4652-645X}}

%\date{\today}

\begin{abstract}
Active matter, comprising many active agents interacting and moving in fluids or more complex environments, is a commonly occurring state of matter in biological and physical systems. By its very nature active matter systems exist in nonequilibrium states. In this paper the active agents are small Janus colloidal particles that use chemical energy provided by chemical reactions occurring on their surfaces for propulsion through a diffusiophoretic mechanism. As a result of interactions among these colloids, either directly or through fluid velocity and concentration fields, they may act collectively to form structures such as dynamic clusters. A general nonequilibrium thermodynamics framework for the description of such systems is presented that accounts for both self-diffusiophoresis and diffusiophoresis due to external concentration gradients, and is consistent with microreversibility. It predicts the existence of a reciprocal effect of diffusiophoresis back onto the reaction rate for the entire collection of colloids in the system, as well as the existence of a clustering instability that leads to nonequilibrium inhomogeneous system states.
\end{abstract}


\maketitle
\section{Introduction}

Active matter is composed of motile entities or agents interacting with each other either directly or through the velocity and concentration fields of the medium in which they move.  Such interactions lead to collective dynamics giving rise to states of matter that may differ from those in equilibrium systems. The study of such collective behavior presents challenges and is currently a topic of considerable scientific interest. Systems with many complex agents can be investigated in different ways. One way is to describe collective dynamics at the macroscale in terms of fields representing the distribution of the agents across the system. These fields are ruled by partial differential equations that are established using general symmetries and experimental observations.  Another approach is to model active matter as being composed of active particles moving in space according to specific rules that are postulated on the basis of empirical considerations.

Both of these approaches have been used to explore the origins and types of collective dynamics that can be found in active matter systems, and research on this topic ranges from studies of simple active particle models, often satisfying minimal rules, to suspensions of more complex active synthetic or biological agents~\cite{V12,RBELS12,A13,CT13,EWG15,SMBL15,BDLRVV16,M16,ZS16,R17,G_20}. The collective behavior in systems where the active agents are chemically-propelled colloids, the subject of this paper, has also been the topic of experimental and theoretical research~\cite{TCPYB12,GPDW14,WDASM15,GTDYC18,TK12,K13,SGR14,PS15,LMC17,HSK17,CK17,RHCK18,HSGK19,S19}.

Systems containing colloidal particles are governed by physico-chemical laws, so that their time evolution can be understood from first principles using statistical-mechanical methods.  This approach was pioneered by Einstein~\cite{E56} and Smoluchowski~\cite{S1915,S1916,S1917} at the beginning of the XXth century and systematically developed since then for passive colloidal particles~\cite{K81,HB83,D96,N13}.  In active matter, the colloidal particles are propelled with energy supplied by the surrounding solution, so that the description should be extended to include the molecular concentrations of fuel and product powering their motion, in addition to the velocity field of the fluid.  Through such an approach, active matter can be described from the scale of a single colloidal motor moving in the surrounding fluid, up to the macroscale where many colloidal motors generate collective motion by interaction.  At the macroscale, collective dynamics is described in terms of the distribution function giving the orientation as well as the position of the colloidal motors.  This statistical-mechanical approach has the advantage that the parameters characterizing active matter at the macroscale can be deduced from the microscopic level of description.  The knowledge of these parameters in terms of the properties of materials composing the colloidal motors and the surrounding solution is fundamental for engineering active systems.

The present paper contributes to the statistical-mechanical and nonequilibrium thermodynamic approaches for active matter systems~\cite{GG10,S12,S_16,JGS_18,FM18,GK17,GK18a,GK19,Speck_19}, and considers systems whose active agents are Janus colloids with catalytic and noncatalytic faces moving by diffusiophoresis generated by chemical reactions taking place on their catalytic faces or caps~\cite{GLA07,OPD17,GK18a}. We start from the calculation of the diffusiophoretic force and torque on a single Janus particle moving in a fluid in the presence of molecular species corresponding to the fuel and the product of the reaction taking place on its catalytic surface.  The concentrations of these molecular species develop gradients under nonequilibrium conditions, and these gradients should be included in the calculation of the force and torque.  The resulting diffusiophoretic force and torque enter the coupled Langevin equations ruling the displacement, rotation, and overall reaction of a single active particle.

Next, the evolution equation is established for the distribution function of the ensemble of active particles in a dilute colloidal solution.  In order to be consistent with microreversibility, the principles of nonequilibrium thermodynamics are used to relate the thermodynamic forces or affinities to the current densities with linear response coefficients satisfying Onsager's reciprocal relations \cite{O31a,O31b,P67,H69,GM84,N79,LL80Part1,C85}.  This method allows us to obtain all the possible couplings compatible with microreversibility, including {\it a priori} unexpected reciprocal effects.  Moreover, this method provides an expression for the entropy production rate density for active matter in agreement with the second law of thermodynamics and including the contribution of the reaction powering activity.  Through this procedure, macroscopic evolution equations are obtained that govern the collective dynamics of colloidal motors coupled to the molecular concentrations of fuel and product.  These equations can be shown to generate the reciprocal effect of diffusiophoresis back onto the reaction rate that has been obtained previously for a single particle \cite{GK17,GK18a}, but now at the macroscale.  Furthermore, pattern formation due to a clustering instability manifests itself under nonequilibrium conditions induced by a bulk reaction replenishing the solution with fuel.

The paper is organized as follows.  Section~\ref{sec:1_motor} is devoted to the dynamics of a single colloidal motor.  The force and torque due to diffusiophoresis are deduced by solving the diffusion equations for the molecular concentrations coupled to the Navier-Stokes equations for the fluid velocity, including the contributions of concentration gradients at large distances from the particle.  These contributions were neglected previously~\cite{GK17,GK18a} and are calculated in detail here.  In Sec.~\ref{sec:N_motors}, the diffusiophoretic force and torque obtained in Sec.~\ref{sec:1_motor} are incorporated into the evolution equation for the distribution function describing the ensemble of colloidal motors, and the entropy production rate density is explicitly obtained.  Two implications of these results are presented in Secs.~\ref{sec:Implications-external} and~\ref{sec:Implications-clustering}.  First, the reciprocal effect due to the diffusiophoretic coupling of an external force and torque back onto the reaction rate is recovered, now at the level of the collective dynamics.  Second, a clustering instability leading to pattern formation is shown to manifest itself. The conclusions of the research are given in Sec.~\ref{sec:Conclusion}. The Appendices provide additional details of the calculations.

\section{Diffusiophoresis and colloidal motors}\label{sec:1_motor}

\begin{figure}[h]
\centerline{\scalebox{0.45}{\includegraphics{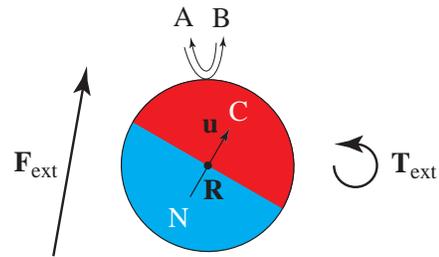}}}
\caption{Schematic representation of a Janus particle with its catalytic (C) and noncatalytic (N) hemispheres where the surface reaction~(\ref{reaction-s}) takes place between fuel A and product B supplied by the solution surrounding the particle.  The particle is also subjected to some external force ${\bf F}_{\rm ext}$ and torque ${\bf T}_{\rm ext}$.  The position of its center of mass is $\bf R$ and $\bf u$ is the unit vector giving its orientation and pointing in the direction of the catalytic hemisphere.}
\label{fig1}
\end{figure}

This section describes the motion of a single spherical Janus colloidal motor of radius $R$ that is propelled by self-diffusiophoresis generated by a reversible reaction
\be
{\rm A} + {\rm C} \underset{\kappa_-}{\stackrel{\kappa_+}{\rightleftharpoons}}  {\rm B} + {\rm C}
\label{reaction-s}
\ee
with rate constants $\kappa_\pm$ taking place on its catalytic surface, as depicted in Fig.~\ref{fig1}.  In this reaction, A is the fuel and B the product, which are present in the solution surrounding the particle.  Moreover, the concentrations of the A and B molecular species are assumed to have gradients ${\bf g}_k$ with $k={\rm A,B}$ at large distances from the particle that also contribute to motion by diffusiophoresis; thus, the motion of the particle is determined by processes in the fluid surrounding the particle.

In order to determine the force and the torque due to diffusiophoresis, as well as the overall reaction rate, the velocity of the fluid and the concentrations of the fuel A and the product B should be obtained by solving the Navier-Stokes equations for the fluid velocity ${\bf v}={\bf v}_{\rm fluid}$ coupled to the advection-diffusion equations for the molecular concentrations $c_k$ with $k={\rm A,B}$:
\bea
&&\rho\left(\partial_t{\bf v} + {\bf v}\cdot\pmb{\nabla}{\bf v}\right) = -\pmb{\nabla}p + \eta \nabla^2{\bf v} \, , \\
&&\pmb{\nabla}\cdot{\bf v} = 0 \, , \\
&&\partial_t \, c_k + {\bf v} \cdot \pmb{\nabla}c_k = D_k {\nabla}^2 c_k \, ,
\eea
where $\rho$ is the constant mass density (the fluid being assumed to be incompressible), $p$~the hydrostatic pressure, $\eta$~the shear viscosity, and $D_k$~the molecular diffusivity of species $k$.

The coupling between the velocity and concentration fields is established with the boundary conditions~\cite{GK18a,GK18b}
\bea
&&{\bf n}\cdot\left({\bf v} -{\bf v}_{\rm solid}\right)_R = 0\, , \label{bc-v-1}\\
&& {\boldsymbol{\mathsf 1}}_{\bot}\cdot\left({\bf v} -{\bf v}_{\rm solid}\right)_{R}
= {\boldsymbol{\mathsf 1}}_{\bot}\cdot\left[ b(\pmb{\nabla}{\bf v})^{\rm S} -\sum_k b_k \pmb{\nabla}c_k\right]_R , \label{bc-v-2} \\
&& D_k \, ({\bf n}\cdot\pmb{\nabla} c_k)_R = -\nu_k \, \, (\kappa_+\, c_{\rm A}-\kappa_-\, c_{\rm B})_R \, , \label{bc-ck}
\eea
where $\bf n$ is the unit vector normal to the solid surface, ${\boldsymbol{\mathsf 1}}_{\bot}\equiv{\boldsymbol{\mathsf 1}}-{\bf n}{\bf n}$, $b$ is the slip length,
$(\pmb{\nabla}{\bf v})^{\rm S} =\left(\pmb{\nabla}{\bf v} + \pmb{\nabla}{\bf v}^{\rm T}\right)$, ${\rm T}$ denotes the transpose, $b_k$ is the diffusiophoretic coefficient of species $k$ coupling the velocity field to the corresponding concentration field because of different interactions between the solid surface with the molecules of different species.  The velocity field inside the solid particle is given by ${\bf v}_{\rm solid}={\bf V} +\pmb{\Omega}\times ({\bf r}-{\bf R})$ in terms of the translational and angular velocities of the particle, respectively denoted by $\bf V$ and $\pmb{\Omega}$.  The last equations are the boundary conditions for the two reacting species $k={\rm A},{\rm B}$, where $\nu_k$ is the stoichiometric coefficient of species $k$ in the reaction ($\nu_{\rm A}=-1$ and $\nu_{\rm B}=+1$), and $\kappa_{\pm}$ are the forward and reverse surface rate constants per unit area.

The velocity field is assumed to vanish at large distances from the particle, so that the entire fluid is at rest.  With the aim of obtaining mean-field equations for a dilute suspension of active particles, we also assume that the concentration fields can have non-vanishing gradients on large spatial scales.  Accordingly, the concentration gradients $(\pmb{\nabla}c_k)_{\infty}={\bf g}_k$ are taken to exist at large distances from the colloidal particle.

We suppose that the diffusiophoretic coefficients take the values $b_k^{\rm c}$ and $b_k^{\rm n}$ on the catalytic and noncatalytic hemispheres, respectively, while the surface rate constants per unit area take positive values $\kappa_{\pm}^{\rm c}$ on the catalytic hemisphere and vanish on the noncatalytic hemisphere, $\kappa_{\pm}^{\rm n}=0$.  Using spherical coordinates $(\theta,\varphi)$ with polar angle $\theta$ defined with respect to the axis of cylindrical symmetry of the Janus particle, we have
\bea
&& b_k(\theta,\varphi) = \sum_{h={\rm c},{\rm n}} b_k^h \, H^h(\theta) \, , \\
&& \kappa_{\pm}(\theta,\varphi) = \sum_{h={\rm c},{\rm n}} \kappa_{\pm}^{h} \, H^h(\theta) \, ,
\eea
where $H^{h}(\theta)$ denotes the Heaviside function such that $H^{h}(\theta)=1$ on hemisphere $h$ and is zero otherwise.  The catalytic hemisphere is taken as $0\le\theta\le\frac{\pi}{2}$ and the noncatalytic hemisphere as $\frac{\pi}{2}<\theta\le\pi$.

The orientation of the Janus particle is described by the unit vector $\bf u$ attached to the axis of cylindrical symmetry of the Janus particle and pointing towards the catalytic hemisphere.  Accordingly, the displacement and the rotation of the particle are ruled by
\be
\frac{d{\bf R}}{dt} = {\bf V} \qquad\mbox{and} \qquad \frac{d{\bf u}}{dt} = \pmb{\Omega}\times{\bf u}
\ee
in terms of the translational and rotational velocities.  These velocities, as well as the number $N$ of reactive events taking place on the particle, are governed by the following coupled Langevin equations~\cite{GK17,GK18a,GK19}:
\begin{eqnarray}
M\frac{d{\bf V}}{dt} &=& -\gamma_{\rm t}\, {\bf V}  + {\bf F}_{\rm d} + {\bf F}_{\rm ext} + {\bf F}_{\rm fl}(t) \, ,
\label{Langevin-eq-trans}\\
{\boldsymbol{\mathsf I}}\cdot\frac{d\pmb{\Omega}}{dt} &=& - \gamma_{\rm r}\,\pmb{\Omega} + {\bf T}_{\rm d} + {\bf T}_{\rm ext} + {\bf T}_{\rm fl}(t) \, ,
\label{Langevin-eq-rot}\\
\frac{dN}{dt} &=& W_{\rm rxn} + W_{\rm d} + W_{\rm fl}(t) \, ,
\end{eqnarray}
where $M$ and ${\boldsymbol{\mathsf I}}$ denote the mass and inertia tensor of the motor, $\gamma_{\rm t}= 6\pi\eta R (1+2b/R)/(1+3b/R)$ is the translational friction coefficient, $\gamma_{\rm r}=8\pi\eta  R^3/(1+3b/R)$ the rotational friction coefficient, ${\bf F}_{\rm d}$ and ${\bf T}_{\rm d}$ the diffusiophoretic force and torque, ${\bf F}_{\rm ext}$ and ${\bf T}_{\rm ext}$ the external force and torque exerted on the particle, while ${\bf F}_{\rm fl}(t)$, ${\bf T}_{\rm fl}(t)$ are the contributions to the force and torque due to thermal fluctuations. The overall net reaction rate is $W_{\rm rxn}$, $W_{\rm d}$ is the reciprocal contribution of diffusiophoresis back onto the reaction rate, and $W_{\rm fl}(t)$ is the fluctuating reaction rate.  If the Janus particle has a magnetic dipole $\mu$ and is subjected to an external magnetic field $\bf B$, then the external torque would be given by ${\bf T}_{\rm ext}=\mu \, {\bf u}\times{\bf B}$.  In the overdamped regime, the coupled Langevin equations are obtained by neglecting the inertial terms in Eqs.~(\ref{Langevin-eq-trans}) and (\ref{Langevin-eq-rot}).

The force and the torque exerted on a spherical particle of radius $R$ in a fluid with shear viscosity $\eta$ and the overall net reaction rate are given by \cite{GK18a}
\bea
&&{\bf F}_{\rm d} = \frac{6\pi\eta R}{1+3b/R} \sum_k  \overline{b_k \, {\boldsymbol{\mathsf 1}}_{\bot}\cdot\pmb{\nabla}c_k}^{\rm s} , \label{Fd} \\
&& {\bf T}_{\rm d} = \frac{12\pi\eta  R}{1+3b/R} \, \sum_k  \overline{b_k \, {\bf r}\times\pmb{\nabla}c_k}^{\rm s} , \label{Td}\\
&& W_{\rm rxn}=4\pi\, R^2 \, \overline{\kappa_+ c_{\rm A}-\kappa_- c_{\rm B}}^{\rm s}  \, ,
\eea
expressed in terms of the surface average
\be
\overline{(\cdot)}^{\rm s} = \frac{1}{4\pi} \int (\cdot)_{r=R}\, d\cos\theta\, d\varphi  \, .
\ee
When writing these equations we have taken into account the possibility that the diffusiophoretic coefficients $b_k$ may be non-uniform on the particle surface.  If molecular diffusion is fast enough so that the concentration fields adopt stationary profiles around the catalytic particle, the diffusiophoretic translational and rotational velocities can be written as follows (see Appendix~\ref{app:FT_motor}):
\bea
&&{\bf V}_{\rm d} = \frac{{\bf F}_{\rm d}}{\gamma_{\rm t}} =  V_{\rm sd} \, {\bf u} + \sum_k \left(\xi_k \, {\boldsymbol{\mathsf 1}}   + \varepsilon_k \, {\boldsymbol{\mathsf Q}}_{\bf u}\right)\cdot {\bf g}_k  \, , \label{Vd}\\
&&\pmb{\Omega}_{\rm d} = \frac{{\bf T}_{\rm d}}{\gamma_{\rm r}} =\sum_k \lambda_k \, {\bf u}\times{\bf g}_k \, ,\label{Omegad}
\eea
where the parameters $\xi_k$, $\varepsilon_k$, and $\lambda_k$ are given in Eqs.~(\ref{eq:v-Omega-parameters2})-(\ref{eq:v-Omega-parameters4}) in terms of the diffusiophoretic coefficients $b_k^h$, the rate constants per unit area $\kappa_{\pm}^{\rm c}$, the slip length $b$, the molecular diffusivities $D_k$, and the geometry of the Janus particle. The $3\times 3$ identity matrix is ${\boldsymbol{\mathsf 1}}$, while
\be\label{Qu}
{\boldsymbol{\mathsf Q}}_{\bf u} \equiv {\bf u}\, {\bf u} - \frac{1}{3} \, {\boldsymbol{\mathsf 1}} \, .
\ee
The self-diffusiophoretic velocity, expressed in terms of the molecular concentrations $\bar{c}_k$ extrapolated to the center of the particle, is
\be\label{Vsd}
V_{\rm sd}= \sum_k \zeta_k \bar{c}_k =\varsigma(\kappa_+^{\rm c}\, \bar{c}_{\rm A}-\kappa_-^{\rm c}\, \bar{c}_{\rm B}),
\ee
since the parameters $\zeta_k$ may be written in the forms $\zeta_{\rm A}=\varsigma \kappa_+^{\rm c}$ and $\zeta_{\rm B}=-\varsigma \kappa_-^{\rm c}$ [see Appendix~\ref{app:FT_motor}, Eq.~(\ref{eq:v-Omega-parameters1})].

In the absence of reaction we recover  the diffusiophoretic velocities given in Refs.~\cite{A89,AP91}:
\begin{eqnarray}
{\bf V}_{\rm d} &=& \sum_k \xi_{k0} \, {\bf g}_k \quad \mbox{with}\quad  \xi_{k0}=\frac{b_k^{\rm c}+b_k^{\rm n}}{2(1+2b/R)}, \\
\pmb{\Omega}_{\rm d} &=& \sum_k \lambda_{k0} \, {\bf u}\times{\bf g}_k \quad  \mbox{with}\quad \lambda_{k0}=\frac{9}{16R}\left( b_k^{\rm c}-b_k^{\rm n}\right)  . \qquad
\end{eqnarray}
Moreover, if the diffusiophoretic coefficients are the same on both hemispheres $b_k^{\rm c}=b_k^{\rm n}$, the angular velocity is equal to zero, $\pmb{\Omega}_{\rm d}=0$.

In the presence of reaction, but without gradients (${\bf g}_k=0$), we have $\kappa_+^{\rm c}\, \bar{c}_{\rm A}\neq\kappa_-^{\rm c}\, \bar{c}_{\rm B}$ and the linear velocity reduces to the contribution of self-diffusiophoresis, ${\bf V}_{\rm d}=V_{\rm sd}{\bf u}$, characterizing the activity of the Janus particle.

The overall reaction rate can be written as follows:
\be\label{W_rxn}
W_{\rm rxn} = k_+\bar{c}_{\rm A}-k_-\bar{c}_{\rm B}+ \varpi\, (k_+{\bf g}_{\rm A}-k_-{\bf g}_{\rm B})\cdot{\bf u} \, ,
\ee
in terms of rate constants $k_{\pm}=\Gamma\kappa_{\pm}^{\rm c}$ and a parameter $\varpi=O(R)$ given in Eq.~(\ref{varpi}). In the absence of the concentration gradients, we recover the expression obtained in Ref.~\cite{GK18a}.  In the presence of the concentration gradients ${\bf g}_k$, there is an extra contribution depending on the direction $\bf u$ of the Janus particle.  However, this last term is normally negligible because we typically have $R\Vert{\bf g}_k\Vert \ll \bar{c}_k$ for micrometric particles and macroscopic gradients of molecular concentrations.

We note that both the self-diffusiophoretic velocity~(\ref{Vsd}) and the leading term of the reaction rate~(\ref{W_rxn}) are proportional to each other.  Their ratio defines the self-diffusiophoretic parameter $\chi$ which was introduced in Refs.~\cite{GK17,GK18a},
\be\label{chi}
\chi \equiv \frac{V_{\rm sd}}{k_+\bar{c}_{\rm A}-k_-\bar{c}_{\rm B}}=\frac{\varsigma}{\Gamma} \, ,
\ee
where the last equality was obtained using $k_{\pm}=\Gamma\kappa_{\pm}^{\rm c}$.

\section{Active solution of colloidal motors}\label{sec:N_motors}

We now show that Onsager's principle of nonequilibrium thermodynamics \cite{O31a,O31b,P67,H69,GM84,N79,LL80Part1,C85} can be used to establish coupled diffusion-reaction equations of motion for active matter that are consistent with microreversibility. According to Onsager's principle, currents are related to thermodynamic forces (or affinities) by
\be
J_{\alpha} = \sum_{\beta} L_{\alpha\beta} \, A^{\beta} \, ,
\label{lin}
\ee
where the linear response coefficients satisfy the Onsager reciprocal relations,
\be\label{Onsager}
L_{\alpha\beta} = L_{\beta\alpha},
\ee
if the affinities are even under time reversal.  The thermodynamic entropy production rate density is given by
\be
\sigma_s = k_{\rm B} \sum_{\alpha} J_{\alpha} \, A^{\alpha} = k_{\rm B} \sum_{\alpha\beta} L_{\alpha\beta} \, A^{\alpha} \, A^{\beta} \ge 0 \, ,
\label{entrprod}
\ee
where $k_{\rm B}$ is Boltzmann's constant.

The system we consider is a dilute solution containing the reactive molecular A and B species, together with colloidal motors~C in an inert solvent S.  The motors are spherical Janus particles and, as described in the previous section, have hemispherical catalytic surfaces where the reaction A~$\rightleftharpoons$~B takes place.  Moreover, we suppose that the solution is at rest, so that the velocity field is equal to zero.  The solution is described at the macroscale in terms of the molecular densities $n_{\rm A}({\bf r},t)$ and $n_{\rm B}({\bf r},t)$, as well as the distribution function of the colloidal motors, $f({\bf r},{\bf u},t)$, where ${\bf r}=(x,y,z)$ is the position and ${\bf u}=(\sin\theta\cos\varphi,\sin\theta\sin\varphi,\cos\theta)$ is the unit vector giving the orientation of the Janus particles (expressed in spherical coordinates in the laboratory frame).  The distribution function is defined as
\be
f({\bf r},{\bf u},t) \equiv \sum_{i=1}^{N_{\rm C}} \delta^3[{\bf r}-{\bf r}_{i}(t)] \, \delta^2[{\bf u}-{\bf u}_i(t)],
\ee
where $\{{\bf r}_i,{\bf u}_i\}_{i=1}^{N_{\rm C}}$ are the positions and orientational unit vectors of the colloidal motors.  For a dilute suspension, the evolution equation of this distribution function can be deduced from the Fokker-Planck equation for the probability that a single colloidal motor is located at the position $\bf r$ with the orientation $\bf u$ \cite{GK17,GK18a,GK19}.  Once, this distribution function is known, we can obtain the successive moments of $\bf u$:
\bea
&&n_{\rm C}({\bf r},t) \equiv \int  f({\bf r},{\bf u},t) \, d^2u \,, \label{nC}\\
&&{\bf p}({\bf r},t) \equiv \int  {\bf u} \, f({\bf r},{\bf u},t) \, d^2u \, , \label{p-vec}\\
&&{\boldsymbol{\mathsf q}}({\bf r},t) \equiv \int {\boldsymbol{\mathsf Q}}_{\bf u}\, f({\bf r},{\bf u},t) \, d^2u \, , \label{q-tensor}\\
&&\qquad\qquad\vdots\nonumber
\eea
where $d^2u=d\cos\theta \, d\varphi$,  $n_{\rm C}$ is the density or concentration of colloidal motors, $\bf p$ is the polarizability or polar order parameter of the colloidal motors, and ${\boldsymbol{\mathsf q}}$ is the traceless order parameter analogous to that for apolar nematic liquid crystals expressed in terms of the tensor~(\ref{Qu}) and, thus, satisfies ${\rm tr}\, {\boldsymbol{\mathsf q}}=0$.

At the macroscale, the reaction is
\be
{\rm A} + {\rm C} \underset{k_-}{\stackrel{k_+}{\rightleftharpoons}}  {\rm B} + {\rm C}
\label{reaction}
\ee
with the rate constants $k_{\pm}$.  If the surface reaction on the Janus colloid was taken into account in by the boundary conditions~(\ref{bc-ck}) in Sec.~\ref{sec:1_motor}, for the colloidal suspension treated here the reaction should be described by a reaction rate density $w$ that is proportional to the distribution function of colloidal motors.

The mean concentrations of molecular species are defined by $n_k=(1-\phi)\bar{c}_k$, where $\phi=4\pi R^3 n_{\rm C}/3$ is the volume fraction of the suspension.  Their corresponding gradients are related to those considered in Sec.~\ref{sec:1_motor} by $\pmb{\nabla}n_k=(1-\phi){\bf g}_k$ for a dilute enough suspension.  The coupled diffusion-reaction equations for the different species take the following forms:
\bea
&& \partial_t n_k + \pmb{\nabla}\cdot\pmb{\jmath}_k= \nu_k\, w  \qquad\quad (k={\rm A,B}) \, , \label{eq-nk}\\
&& \partial_t f +\pmb{\nabla}\cdot\left({\bf V}  f -D_{\rm t}\pmb{\nabla}f\right) = D_{\rm r}\, \hat{\cal L}_{\rm r} f \, , \label{master}
\eea
where $\pmb{\jmath}_k$ are the molecular current densities, $\bf V$ is the total drift velocity obtained by adding the drift velocity due to the external force ${\bf V}_{\rm ext}={\bf F}_{\rm ext}/\gamma_{\rm t}$ to the diffusiophoretic velocity~(\ref{Vd}) giving
\be
{\bf V} = V_{\rm sd} \, {\bf u} + \sum_{k} \left( \xi_k \, {\boldsymbol{\mathsf 1}}+ \varepsilon_k \, {\boldsymbol{\mathsf Q}}_{\bf u}\right)\cdot \pmb{\nabla} n_k + \beta D_{\rm t} {\bf F}_{\rm ext} \, , \label{Vd2}
\ee
with the self-diffusiophoretic velocity [Eq.~(\ref{Vsd})]
\be
V_{\rm sd}= \sum_{k} \zeta_k \,  n_k = \varsigma (\kappa_+^{\rm c} n_{\rm A} -\kappa_-^{\rm c} n_{\rm B}) \, ,
\ee
now expressed in terms of the mean concentrations $n_k$.  In Eq.~(\ref{master}), $D_{\rm t}$ is an effective translational diffusion coefficient related to the effective translational friction coefficient by Einstein's formula $D_{\rm t}= k_{\rm B}T/\gamma_{\rm t}$ and $D_{\rm r}$ is an effective rotational diffusion coefficient related to the effective rotational friction coefficient by $D_{\rm r}= k_{\rm B}T/\gamma_{\rm r}$.  Since the shear viscosity increases as $\eta\simeq\eta^{(0)}(1+2.5\phi)$ with the volume fraction $\phi$ of the suspension \cite{E56,HB83}, both friction coefficients $\gamma_{\rm t}$ and $\gamma_{\rm r}$ also increase, and the diffusion coefficients decrease.  In particular, it is known that $D_{\rm t}\simeq D_{\rm t}^{(0)} (1-2.1 \phi)$~\cite{HB83}.  A similar dependence on the volume fraction $\phi$ is expected for the parameters $\varsigma$, $\xi_k$, $\varepsilon_k$, and $\lambda_k$ given in Appendix~\ref{app:FT_motor}, since these parameters are proportional to the diffusiophoretic coefficients $b_k^h$ that are known to be inversely proportional to shear viscosity, $b_k^h\propto \eta^{-1}$ \cite{GK18b,A89,AP91}.  The effects of this dependence would manifest themselves if the colloidal suspension became dense enough.  Here, such effects are assumed to play a negligible role.

The Janus particles have a spherical shape so that their random rotational and translational motions are decoupled.  In this case the rotational diffusion operator is given by
\begin{eqnarray}
\hat{\cal L}_{\rm r} f &=& \frac{1}{\sin\theta}\, \partial_{\theta} \left[ \sin\theta \,
{\rm e}^{-\beta U_{\rm r}} \partial_{\theta}\left( {\rm e}^{\beta U_{\rm r}} f \right) \right] \nonumber \\
&&+ \frac{1}{\sin^2\theta}\, \partial_{\varphi} \left[  {\rm e}^{-\beta U_{\rm r}} \partial_{\varphi}\left( {\rm e}^{\beta U_{\rm r}} f \right) \right] ,
\label{Lr}
\end{eqnarray}
expressed in terms of the inverse temperature $\beta=(k_{\rm B}T)^{-1}$ and the rotational energy associated with the torque exerted by an external magnetic field ${\bf B}$ on some magnetic dipole $\mu$ of the particle~\cite{LL80Part1} and that due to the diffusiophoretic effect:
\be
U_{\rm r} = -\mu \, {\bf B} \cdot{\bf u} -\gamma_{\rm r} \Big(\sum_{k} \lambda_k \,  \pmb{\nabla}n_k \Big)\cdot{\bf u} \, . \label{pot}
\ee

The distribution function $f({\bf r},{\bf u})$ for colloidal Janus particles is defined in the five-dimensional space $(x,y,z,\theta,\varphi)$.  For the rotational degrees of freedom we have
\begin{eqnarray}\label{metric}
d{\bf u}^2 &=&d\theta^2 + \sin^2\theta \, d\varphi^2 \\
&=& g_{ij}\, dq^{i} \, dq^{j} \quad\mbox{with} \quad
(g_{ij}) = \left(
\begin{array}{cc}
1 &  0 \\
0 & \sin^2\theta
\end{array}
\right) .\nonumber
\end{eqnarray}
The scalar product between a pair of rotational vectors ${\bf a}_{\rm r}, {\bf b}_{\rm r} \in{\mathbb R}^2$ is given by ${\bf a}_{\rm r}\bullet{\bf b}_{\rm r}=\sum_{i,j=\theta,\varphi} g_{ij} a_{\rm r}^i b_{\rm r}^j$ and the scalar product of such a vector with itself is denoted ${\bf a}_{\rm r}^2={\bf a}_{\rm r}\bullet{\bf a}_{\rm r}$.  In spherical coordinates, the rotational gradient and divergence are given, respectively, by \cite{B65}
\be\label{grad_r}
{\rm grad}_{\rm r}X=\left(
\begin{array}{c}
\partial_{\theta} X \\
\frac{1}{\sin^2\theta}\, \partial_{\varphi} X
\end{array}\right),
\ee
\be\label{div_r}
{\rm div}_{\rm r} \,\bm{X}_{\rm r}= \frac{1}{\sin\theta} \, \partial_{\theta}\left( X_{\rm r}^{\theta}\, \sin\theta\right) + \partial_{\varphi} \, X_{\rm r}^{\varphi}  \, .
\ee
In the five-dimensional space the gradient is given by
\be
{\rm grad}\, X =
\left(\begin{array}{c}
\pmb{\nabla} X \\ {\rm grad}_{\rm r} X
\end{array}\right) \quad\mbox{with}\quad \pmb{\nabla} X= \left(\begin{array}{c}
\partial_x X \\ \partial_y X \\ \partial_z X
\end{array}\right),
\ee
and the divergence of a five-dimensional vector $\bm{X}=(\bm{X}_{\rm t}, \bm{X}_{\rm r})^{\rm T}$, is
\be
{\rm div} \bm{X}=\pmb{\nabla}\cdot\bm{X}_{\rm t} + {\rm div}_{\rm r} \, \bm{X}_{\rm r}.
\ee

Using these notations, Eq.~(\ref{master}) can be written in the form of a local conservation law involving the five-dimensional current density, $\pmb{\cal J}_{\rm C} = (\pmb{\jmath}_{\rm t}, \pmb{\jmath}_{\rm r})^{\rm T}$, as
\be
\partial_t f + {\rm div}\, \pmb{\cal J}_{\rm C} = 0 \quad\mbox{or} \quad \partial_t f +\pmb{\nabla}\cdot\pmb{\jmath}_{\rm t} + {\rm div}_{\rm r} \, \pmb{\jmath}_{\rm r} =0 \, ,
\label{master2}
\ee
with translational current density
\begin{eqnarray}\label{JCt}
&&\pmb{\jmath}_{\rm t}= {\bf V} f -D_{\rm t}\pmb{\nabla}f = f\, V_{\rm sd} \, {\bf u} \\
&& \; + f  \sum_{k} \left(\xi_k\, {\boldsymbol{\mathsf 1}} + \varepsilon_k \, {\boldsymbol{\mathsf Q}}_{\bf u}\right)\cdot\pmb{\nabla} n_k
-D_{\rm t} \left(\pmb{\nabla} f  +f\beta \pmb{\nabla} U_{\rm t}\right) \, , \label{j_t} \nonumber
\end{eqnarray}
rotational current density
\begin{eqnarray}\label{JCr}
\pmb{\jmath}_{\rm r}  &=&- D_{\rm r} \, {\rm e}^{-\beta U_{\rm r}} \, {\rm grad}_{\rm r}({\rm e}^{\beta U_{\rm r}}   f ) \nonumber \\ &=& f\big(\sum_{k}\lambda_k\, \pmb{\nabla} n_k\big)\cdot {\rm grad}_{\rm r} {\bf u}\nonumber \\
 &&-D_{\rm r} \left({\rm grad}_{\rm r} f - f\, \beta\, \mu \, {\bf B}\cdot{\rm grad}_{\rm r} {\bf u}\right),
\label{j_r}
\end{eqnarray}
and their translational and rotational divergences, $\pmb{\nabla}\cdot\pmb{\jmath}_{\rm t}$ and ${\rm div}_{\rm r} \, \pmb{\jmath}_{\rm r}=-D_{\rm r} \hat{\cal L}_{\rm r} f$, where $\hat{\cal L}_{\rm r}$ is the operator~(\ref{Lr}).

Furthermore, we suppose that the system is isothermal and isobaric and the solution is dilute in the species A, B, and C. The appropriate thermodynamic potential is thus Gibbs' free energy given by
\bea\label{free_energy_G}
&G&= \int d^3r \bigg\{ n_{\rm S}  \psi_{\rm S} + \sum_{k={\rm A,B}} \left(n_k \psi_k + n_k  k_{\rm B}T  \ln\frac{n_k}{{\rm e} n_{\rm S}} \right) \nonumber\\
&\; +& \int d^2u \Big[ f  \psi_{\rm C} + f k_{\rm B}T  \ln \frac{f}{4\pi{\rm e} n_{\rm S}} + f  U_{\rm t}({\bf r}) - f  \mu {\bf B}\cdot{\bf u} \Big] \bigg\},\nonumber \\
\eea
where $U_{\rm t}({\bf r})=-{\bf F}_{\rm ext}\cdot{\bf r}$ the translational potential energy due to the external force ${\bf F}_{\rm ext}$. Thus, we deduce the following chemical potentials,
\bea
&& \mu_{\rm S} = \frac{\delta G}{\delta n_{\rm S}} = \psi_{\rm S} - \frac{k_{\rm B}T}{n_{\rm S}} \, (n_{\rm A}+n_{\rm B}+n_{\rm C}) \, , \label{mu-S} \\
&& \mu_k = \frac{\delta G}{\delta n_k} = \psi_k + k_{\rm B}T \ln\frac{n_k}{n_{\rm S}} \quad\quad (k={\rm A,B}) \, , \label{mu-k}\\
&& \mu_{\rm C} = \frac{\delta G}{\delta f} = \psi_{\rm C} + k_{\rm B}T \ln\frac{f}{4\pi n_{\rm S}} + U_{\rm t}({\bf r}) - \mu {\bf B}\cdot{\bf u}.\nonumber \\ \label{mu-C}
\eea
Here $\psi_k=\mu_k^0+k_{\rm B}T\ln(n_{\rm S}/n^0)$, where $\mu_k^0$ is the standard chemical potential of species $k$ and $n^0=1~{\rm mole/liter}$ is the standard concentration.
Since the solution is dilute, we have taken the solvent density $n_{\rm S}$ to be essentially uniform in space and constant in time.

Next we use the principles of nonequilibrium thermodynamics in order to express the current densities in terms of the affinities or thermodynamic forces.  For the reaction~(\ref{reaction}), the affinity is given by
\be\label{Arxn}
A_{\rm rxn} = \frac{1}{k_{\rm B}T} \, \left(\mu_{\rm A}-\mu_{\rm B}\right) =  \ln \frac{ k_+ n_{\rm A}}{ k_- n_{\rm B}} \, .
\ee
At chemical equilibrium, we have $A_{\rm rxn}=0$, $w=0$, and $k_+ n_{\rm A,eq} =k_- n_{\rm B,eq}$.  In the linear regime close to equilibrium where $\delta n_k = n_k-n_{k,{\rm eq}}$, the chemical affinity~(\ref{Arxn}) can be approximated as
\be\label{Arxn2}
A_{\rm rxn} = \frac{\delta n_{\rm A}}{n_{\rm A,eq}} -  \frac{\delta n_{\rm B}}{n_{\rm B,eq}} =  \frac{1}{D_{\rm rxn}} \left( k_+\delta n_{\rm A}-k_-\delta n_{\rm B}\right) ,
\ee
where
\be
D_{\rm rxn} \equiv \frac{1}{2}\left( k_+ \, n_{\rm A} + k_-\, n_{\rm B} \right)_{\rm eq}
\ee
is the diffusivity of the reaction taking place on the colloidal motors~\cite{GK17,GK18a}.  For the diffusion processes of species $k$, the affinity associated with the current density $\pmb{\jmath}_{k}$ is given by ${\bf A}_{k}= -{\rm grad}\frac{\mu_k}{k_{\rm B}T}$ in terms of the chemical potential $\mu_k$.  For molecular species, the gradient is tridimensional in Euclidian space, so that ${\bf A}_{k}= -\pmb{\nabla}\frac{\mu_k}{k_{\rm B}T}=-n_k^{-1}\pmb{\nabla}n_k$.
For the colloid with chemical potential~(\ref{mu-C}), we have that
\be
{\rm grad}\, \frac{\mu_{\rm C}}{k_{\rm B}T}=
\frac{1}{f} \left(\begin{array}{c}
\pmb{\nabla}f +f\beta\pmb{\nabla} U_{\rm t} \\
{\rm grad}_{\rm r} f-f\beta\mu\, {\bf B}\cdot{\rm grad}_{\rm r}{\bf u}
\end{array}\right) ,
\ee
if the magnetic field $\bf B$ is uniform.
In this five-dimensional space, the associated current density $\pmb{\cal J}_{\rm C}$ can therefore be written in the following form:
\begin{eqnarray}
\pmb{\cal J}_{\rm C} &=& f \left(
\begin{array}{c}
 V_{\rm sd} \, {\bf u} \\
{\bf 0}
\end{array}
\right) +
 f \sum_{k={\rm A,B}} \left(
\begin{array}{c}
 \xi_k\, {\boldsymbol{\mathsf 1}} + \varepsilon_k \, {\boldsymbol{\mathsf Q}}_{\bf u}\\
\lambda_k\, {\rm grad}_{\rm r} {\bf u}
\end{array}
\right) \cdot \pmb{\nabla} n_k \nonumber \\
&&- f  \left(
\begin{array}{cc}
 D_{\rm t} {\boldsymbol{\mathsf 1}} &  {\boldsymbol{\mathsf 0}}   \\
 {\boldsymbol{\mathsf 0}} &  D_{\rm r}  {\boldsymbol{\mathsf 1}}_{\rm r}
 \end{array}
\right) \cdot {\rm grad} \, \frac{\mu_{\rm C}}{k_{\rm B}T} \, ,
\label{JC2}
\end{eqnarray}
where ${\boldsymbol{\mathsf 1}}_{\rm r}$ is the $2\times 2$ identity matrix.
In this form, we see that the first term is related to the reaction affinity since the self-diffusiophoretic velocity can be written as $V_{\rm sd} = \chi \, D_{\rm rxn} \, A_{\rm rxn}$.
The next two terms can be related to the affinities of molecular species, and the last term to the affinity of the colloidal species.

According to the Curie principle, there is no coupling between processes with different tensorial character.  However, the Janus particles have a director given by the unit vector $\bf u$ and we have adopted a description in terms of the distribution function $f({\bf r},{\bf u},t)$ for the Janus particles.  Consequently, it is possible that a vectorial process such as diffusion may be coupled to a scalar process such as reaction if it is polarized by the unit vector $\bf u$.  If we introduce the densities ${\mathscr N}_{\rm C} = f \, \Delta^2u$ for Janus particles having their orientation $\bf u$ in cells of size $\Delta^2 u$, along with the associated current densities,
\be
{\bf J}_{\rm C} = \pmb{\cal J}_{\rm C} \, \Delta^2u \, ,
\ee
we may write a general coupling~(\ref{lin}) of the following form:
\be\label{J=LA}
\left(\begin{array}{c}
w \\
\pmb{\jmath}_{\rm A}\\
\pmb{\jmath}_{\rm B}\\
{\bf J}_{\rm C}
\end{array}
\right)
=
\left(\begin{array}{cccc}
L_{\rm rr} & {\bf L}_{\rm rA} & {\bf L}_{\rm rB} &  {\bf L}_{\rm rC}  \\
{\bf L}_{\rm Ar} & {\boldsymbol{\mathsf L}}_{\rm AA} & {\boldsymbol{\mathsf L}}_{\rm AB}& {\boldsymbol{\mathsf L}}_{\rm AC} \\
{\bf L}_{\rm Br} & {\boldsymbol{\mathsf L}}_{\rm BA} & {\boldsymbol{\mathsf L}}_{\rm BB}& {\boldsymbol{\mathsf L}}_{\rm BC} \\
{\bf L}_{\rm Cr} & {\boldsymbol{\mathsf L}}_{\rm CA} & {\boldsymbol{\mathsf L}}_{\rm CB}& {\boldsymbol{\mathsf L}}_{\rm CC}
\end{array}
\right)\cdot
\left(\begin{array}{c}
A_{\rm rxn} \\
-\pmb{\nabla}\frac{\mu_{\rm A}}{k_{\rm B}T}\\
-\pmb{\nabla}\frac{\mu_{\rm B}}{k_{\rm B}T}\\
-{\rm grad}\, \frac{\mu_{\rm C}}{k_{\rm B}T}
\end{array}
\right) \, ,
\ee
up to possible nonlinear contributions that may be required in order for the reaction rate to obey the mass-action law.  In Eq.~(\ref{J=LA}), we have that $L_{\rm rr}$ is $1\times 1$, ${\bf L}_{{\rm r}k}$ $1\times 3$, ${\bf L}_{\rm rC}$ $1\times 5$, ${\bf L}_{k{\rm r}}$ $3\times 1$, ${\boldsymbol{\mathsf L}}_{kl}$ $3\times 3$, ${\boldsymbol{\mathsf L}}_{k{\rm C}}$ $3\times 5$, ${\bf L}_{\rm Cr}$ $5\times 1$, ${\boldsymbol{\mathsf L}}_{{\rm C}k}$ $5\times 3$, and ${\boldsymbol{\mathsf L}}_{\rm CC}$ $5\times 5$ (for $k,l={\rm A,B}$).

According to Onsager's reciprocal relations~(\ref{Onsager}), the linear response coefficients should obey
\be
{\bf L}_{{\rm r}k} ={\bf L}_{k{\rm r}}^{\rm T} \, , \qquad {\bf L}_{\rm rC} ={\bf L}_{\rm Cr}^{\rm T} \, ,
\ee
\be
{\boldsymbol{\mathsf L}}_{kl}={\boldsymbol{\mathsf L}}_{lk}^{\rm T} \, , \qquad
{\boldsymbol{\mathsf L}}_{\rm CC}={\boldsymbol{\mathsf L}}_{\rm CC}^{\rm T} \, , \qquad
{\boldsymbol{\mathsf L}}_{k{\rm C}}={\boldsymbol{\mathsf L}}_{{\rm C}k}^{\rm T} \, ,
\ee
for $k={\rm A,B}$ and where ${\rm T}$ again denotes the transpose.

We assume that the molecular species A and B undergo Fickian diffusion without cross-diffusion, so that
\be\label{Fick}
{\boldsymbol{\mathsf L}}_{kl}= D_k \, n_k \, \delta_{kl} \,  {\boldsymbol{\mathsf 1}} \, ,
\ee
and that the reaction rate does not depend on the gradients $\pmb{\nabla}n_{\rm A}$ or $\pmb{\nabla}n_{\rm B}$, whereupon
\be\label{Lrk=0}
{\bf L}_{\rm rA} = {\bf L}_{\rm rB} = 0 \, .
\ee
This last assumption consists in neglecting the terms with the coefficient $\varpi$ in Eq.~(\ref{W_rxn}), which is usually justified as mentioned in Sec.~\ref{sec:1_motor}.

The scalar coefficient associated with the reaction can be identified as
\be
L_{\rm rr} = D_{\rm rxn} \, n_{\rm C} \, ,
\ee
and the linear response coefficients ${\bf L}_{\rm Cr}$, ${\boldsymbol{\mathsf L}}_{{\rm C}k}$,
and $ {\boldsymbol{\mathsf L}}_{\rm CC}$ in Eq.~(\ref{J=LA}) can be determined using the current density~(\ref{JC2}), as described in App.~\ref{app:Onsager}.  As a consequence of Onsager's reciprocal relations, we can conclude that the reaction rate and the current densities should be given by
\bea
&& w = D_{\rm rxn} n_{\rm C} A_{\rm rxn} - \chi  D_{\rm rxn} \int {\bf u}\cdot\left(\pmb{\nabla}f+f\beta\pmb{\nabla}U_{\rm t}\right)  d^2u  , \label{w}\\
&& \pmb{\jmath}_k = -D_k\, \pmb{\nabla} n_k + n_k \int\Big[\left(\xi_k {\boldsymbol{\mathsf 1}} + \varepsilon_k {\boldsymbol{\mathsf Q}}_{\bf u}\right)\cdot\left(\pmb{\nabla}f+f\beta\pmb{\nabla}U_{\rm t}\right) \nonumber \\
 &&\quad + \lambda_k ({\rm grad}_{\rm r} {\bf u})\bullet\left({\rm grad}_{\rm r}  f - f \beta \mu \, {\bf B}\cdot {\rm grad}_{\rm r} {\bf u}\right)\Big]   d^2u\, .  \label{Jk}
\eea
In Eq.~(\ref{w}), the second term describes the reciprocal effects of diffusiophoresis back onto reaction.  The second term in Eq.~(\ref{Jk}) is due to cross-diffusion between the molecular and colloidal species due to diffusiophoresis.  We see that the linear response coefficients depend on the unit vector $\bf u$ in a manner similar to that already shown in Refs.~\cite{GK17,GK18a}.

With respect to standard expressions, the terms involving the integral $\int d^2u$ in Eq.~(\ref{Jk}) are required in order to satisfy Onsager's reciprocal relations and for these quantities to be compatible with microreversibility.  However, these extra terms can be shown to be negligible, although the reciprocal terms are not negligible in Eqs.~(\ref{JCt}) and (\ref{JCr}).  In order to show that the extra terms are negligible, we suppose that the self-diffusiophoretic and diffusiophoretic velocities take the typical value $V_{\rm sd} \sim V_{\rm d} \sim 10\,  \mu{\rm m/s}$ \cite{W13}.  According to Ref.~\cite{AP91}, the molecular gradients used in experiments of diffusiophoresis are of the order of $\Vert\pmb{\nabla} n_k \Vert \sim 10^5 \, {\rm mol/m}^4$,  so that diffusiophoretic parameters have the value $\xi_k, \varepsilon_k \sim 10^{-10} \, {\rm m}^5\, {\rm s}^{-1} \, {\rm mol}^{-1}$.  Moreover, we have $\lambda_k\sim\xi_k/R$, but since $\Vert{\rm grad}_{\rm r}f \Vert \sim R \Vert\pmb{\nabla} f\Vert$, the effect of the coefficients $\lambda_k$ is again of the same order of magnitude as $\xi_k$ and $\varepsilon_k$. Molecular diffusivities typically have the value $D_k \sim 10^{-9} \, {\rm m}^2/{\rm s}$, while the translational diffusion coefficient of a micrometric colloidal particle is of the order of $D_{\rm t} \sim 10^{-13} \, {\rm m}^2/{\rm s}$.  The molecular concentrations used in experiments on self-diffusiophoresis are about $n_k \sim 10^3 \, {\rm mol}/{\rm m}^3$, while the density of micrometric colloidal particles is approximately $n_{\rm C} \sim 10^{18} \, {\rm m}^{-3} \sim 10^{-6} \, {\rm mol}/{\rm m}^3$, or lower.  If we assume that the molecular and colloidal gradients take comparable values $\Vert\pmb{\nabla} n_k\Vert/n_k \sim \Vert\pmb{\nabla} f\Vert/f$, the ratio between the extra term and the standard molecular diffusion term in Eq.~(\ref{Jk}) is given by
\be
\frac{n_k\, \xi_k\, \Vert\pmb{\nabla} f\Vert}{D_k\, \Vert\pmb{\nabla} n_k\Vert} \sim
\frac{\xi_k\, f}{D_k} \sim 10^{-8} \, ,
\ee
which shows that the second term in Eq.~(\ref{Jk}) is negligible.  Accordingly, the standard Fickian expressions $\pmb{\jmath}_k \simeq -D_k\, \pmb{\nabla} n_k$ are very well justified for the molecular current densities.  In the presence of colloidal motors, the expressions compatible with microreversibility are nevertheless given by Eqs.~(\ref{w}) and~(\ref{Jk}).  In contrast, the terms associated with the diffusiophoretic parameters  in the colloidal current density~(\ref{JC2}) have effects that are not negligible.

The conclusion from these considerations is that active matter can be described as generalized diffusion-reaction processes in complete compatibility with microreversibility and Onsager's reciprocal relations.  In this way, the program of nonequilibrium thermodynamics is complete and application of Eq.~(\ref{entrprod}) gives the following expression for the thermodynamic entropy production rate density:
\begin{widetext}
\bea
k_{\rm B}^{-1} \, \sigma_s &=& D_{\rm rxn}\, n_{\rm C} \, A_{\rm rxn}^2 + \sum_{k={\rm A,B}} D_k \frac{(\pmb{\nabla}n_k)^2}{n_k} - 2\, \chi \,D_{\rm rxn} \,A_{\rm rxn} \int {\bf u}\cdot\left(\pmb{\nabla}f+f\beta\pmb{\nabla}U_{\rm t}\right)  d^2u \nonumber\\
&-& 2 \int \biggl[ \,\sum_{k={\rm A,B}} \pmb{\nabla} n_k\cdot\left(\xi_k\, {\boldsymbol{\mathsf 1}} + \varepsilon_k \, {\boldsymbol{\mathsf Q}}_{\bf u}\right)\biggr]\cdot\left(\pmb{\nabla}f+f\beta\pmb{\nabla}U_{\rm t}\right)  d^2u  \nonumber\\
&-& \int \biggl[\biggl(\,\sum_{k={\rm A,B}}\lambda_k\, \pmb{\nabla} n_k\biggr)\cdot\left({\rm grad}_{\rm r} {\bf u}\right)\biggr] \bullet\left({\rm grad}_{\rm r}  f - f \beta \mu \, {\bf B}\cdot {\rm grad}_{\rm r} {\bf u}\right)  d^2u  \nonumber\\
&+& D_{\rm t} \int \frac{1}{f} \left(\pmb{\nabla}f+f\beta\pmb{\nabla}U_{\rm t}\right)^2  d^2u + D_{\rm r}\, \int \frac{1}{f} \left({\rm grad}_{\rm r}  f - f \beta \mu \, {\bf B}\cdot {\rm grad}_{\rm r} {\bf u}\right)^2 d^2u \ \ge 0 \, .
\eea
\end{widetext}
The second law is satisfied if $D_{\rm t} \gg 4\pi \chi^2 D_{\rm rxn}>0$, $D_kD_{\rm t} \gg 4\pi \xi_k^2>0$,  $D_kD_{\rm t} \gg 4\pi \varepsilon_k^2>0$, $D_kD_{\rm t} \gg 4\pi \lambda_k^2>0$, and $D_{\rm r}>0$, which is expected.

The results derived in this section provide the basis for the analysis of collective effects in suspensions of active Janus particles. In the next sections, we describe two collective phenomena that emerge from this theoretical framework: the effect of an external force and torque on the reaction rate, and a clustering instability.

\section{Effect of external force and torque}
\label{sec:Implications-external}

Using a thermodynamic formulation that is consistent with microreversibility, we showed earlier~\cite{GK17,GK18a,GK19,HSGK18} how the application of an external force and torque on a single colloidal motor can change the reaction rate on its surface and even lead to a net production of fuel rather than product. Now we show how these considerations can be extended to a suspension of such motors.

We suppose that the colloidal motors are subjected to an external force ${\bf F}_{\rm ext}=F_{\rm ext} {\bf 1}_z$ and an external torque induced by an external magnetic field ${\bf B}=B{\bf 1}_z$ exerted on the magnetic moment $\mu$ of the colloidal particles, both oriented in the $z$-direction.  If $\beta\mu B$ is large enough, the distribution function is given by
\be
f({\bf r},{\bf u},t) = n({\bf r},t) \, \frac{\beta\mu B}{4\pi \, \sinh \beta\mu B} \, \exp(\beta\mu B\cos\theta) \, ,
\ee
so that ${\bf p}({\bf r},t) = {\bf 1}_z \langle u_z\rangle\, n_{\rm C}({\bf r},t)$ with $\langle u_z\rangle=\coth\beta\mu B-(\beta\mu B)^{-1}$.  Moreover, the terms with the coefficients $\xi_k$, $\varepsilon_k$, and $\lambda_k$ are assumed to be negligible in Eq.~(\ref{master}).  If the concentrations are uniform in the $x$- and $y$-directions, the process is ruled by the following equation:
\begin{eqnarray}\label{eq-nC-Fext}
&&\partial_t n_{\rm C} + \partial_z \big[ \chi \langle u_z\rangle (k_+ n_{\rm A}-k_- n_{\rm B}) n_{\rm C} \\
&& \qquad \qquad - D_{\rm t} (\partial_z n_{\rm C}-\beta F_{\rm ext} n_{\rm C}) \big] = 0 ,\nonumber
\end{eqnarray}
obtained by integrating Eq.~(\ref{master}) over the orientation $\bf u$.  This equation for $n_{\rm C}$ is coupled to Eq.~(\ref{eq-nk}) with Fickian molecular current densities $\pmb{\jmath}_k \simeq -D_k\, \pmb{\nabla} n_k$ and the local reaction rate
\be\label{local_rate-Fext}
w = (k_+ n_{\rm A}-k_- n_{\rm B}) n_{\rm C} - \chi \langle u_z\rangle D_{\rm rxn} (\partial_z n_{\rm C}-\beta F_{\rm ext} n_{\rm C})
\ee
given by Eq.~(\ref{w}) predicted by Onsager's reciprocal relations.
Defining the mean value of the $z$-coordinate for the colloidal motors as
\be\label{dzdt}
\langle z\rangle \equiv \frac{\int z\, n_{\rm C} \, d^3r}{\int n_{\rm C} \, d^3r} \, ,
\ee
and using Eq.~(\ref{eq-nC-Fext}), we obtain the evolution equation,
\be
\frac{d\langle z\rangle}{dt} = \chi\langle u_z\rangle w_{\rm rxn} + \beta D_{\rm t} F_{\rm ext}
\ee
with mean reaction rate,
\be \label{eq:mean-rate}
w_{\rm rxn} \equiv k_+ \frac{\int n_{\rm A}\, n_{\rm C} \, d^3r}{\int n_{\rm C} \, d^3r}
- k_- \frac{\int n_{\rm B}\, n_{\rm C} \, d^3r}{\int n_{\rm C} \, d^3r} \, .
\ee
Furthermore, integrating Eq.~(\ref{eq-nk}) with $k={\rm B}$ over the position $\bf r$ with the local rate~(\ref{local_rate-Fext}), we get the total reaction rate
\be\label{dNdt}
\frac{dN_{\rm rxn}}{dt} = \frac{dN_{\rm B}}{dt} = - \frac{dN_{\rm A}}{dt} = N_{\rm C} \left( w_{\rm rxn} + \chi \langle u_z\rangle D_{\rm rxn} \beta F_{\rm ext} \right) ,
\ee
where $N_{\rm C}=\int n_{\rm C} \, d^3r$ is the total number of colloidal motors in the suspension. Equations~(\ref{dzdt}) and (\ref{dNdt}) have precisely the same structure as for a single colloidal motor.  However, one should note that the mean reaction rate in Eq.~(\ref{eq:mean-rate}) contains spatial correlations between the solute and colloid concentration fields. Given the structure of the equations, the results obtained in Refs.~\cite{GK17,GK18a} also apply here.  In particular, there exists a regime where the entire ensemble of colloidal motors is propelled and carries out work against the external force by consuming fuel.  In addition, there is also a regime where fuel is synthesized if the external force that opposes motion is sufficiently large to reverse the reaction ${\rm A}\to{\rm B}$.  The efficiencies of these processes are given by the same expressions as in Refs.~\cite{GK17,GK18a}.

\section{Clustering instability and pattern formation}\label{sec:Implications-clustering}

The equations of motion developed in Sec.~\ref{sec:N_motors} that describe a dilute suspension of colloidal motors moving in a dilute solution of fuel A and product B molecular species will be shown in this section to lead to a clustering instability.  This instability can be described by the mean-field equations obtained above for the concentrations of the molecular species and the distribution function of the colloidal motors. A number of other mean-field descriptions that predict instabilities and the formation of various clustering states of collections of diffusiophoretic colloidal particles have appeared in literature~\cite{SBML14,SGR14,SMBL15,PS15,LMC17}, and use techniques involving coupled moment equations similar to those adopted in this section.

The equation for the colloidal motors is coupled to the reaction-diffusion equations for the molecular species A and B, accounting for the fact that the reaction ${\rm A}\rightleftharpoons {\rm B}$ occurs both at the surface of the catalytic hemisphere of the colloids and in the bulk:
\be
\partial_t n_k = D_k\nabla^2 n_k + \nu_k \, w_{\rm tot}  \, , \label{eq-nk-2}
\ee
where the total reaction rate density is given by
\be
w_{\rm tot} = (k_+n_{\rm A}-k_-n_{\rm B})\, n_{\rm C}+ k_{+2} n_{\rm A}-k_{-2}n_{\rm B} \, .
\label{rate}
\ee
The system is driven out of equilibrium if $k_+/k_- \ne k_{+2}/k_{-2}$~\cite{HSGK18}.

If the second moment~(\ref{q-tensor}) as well as higher moments are assumed to be negligible,
the evolution equations for the density of colloidal motors~(\ref{nC}) and the polarizability~(\ref{p-vec}) are given by
\be
\partial_t n_{\rm C} +\pmb{\nabla}\cdot\left(n_{\rm C}\sum_k \xi_k  \pmb{\nabla} n_k  + V_{\rm sd} \, {\bf p} \right) = D_{\rm t}\nabla^2 n_{\rm C}  , \quad\label{eq-nC-2}
\ee
\bea
\partial_t {\bf p} &+& \pmb{\nabla}\cdot\left({\bf p} \, \sum_k \xi_k \, \pmb{\nabla} n_k\right)  + \frac{1}{3}\, \pmb{\nabla}\left(V_{\rm sd} \, n_{\rm C}\right)  \nonumber \\
&+&\frac{1}{5}\, {\boldsymbol{\mathsf\Delta}}:\pmb{\nabla}\left(\sum_k \varepsilon_k\,  \pmb{\nabla} n_k\, {\bf p}\right) \nonumber\\
&=& D_{\rm t} \nabla^2 {\bf p}-2D_{\rm r} {\bf p} + \frac{2}{3}\, n_{\rm C} \sum_k\lambda_k\pmb{\nabla} n_k \, , \label{eq-p-2}
\eea
in terms of the fourth-order tensor ${\boldsymbol{\mathsf\Delta}}$ with the following components:
$\Delta_{ijmn} = \delta_{ij}\, \delta_{mn} +\delta_{im}\, \delta_{jn} -\frac{2}{3} \, \delta_{in}\, \delta_{jm}$.

If $D_{\rm r}$ is large enough so that $2D_{\rm r} {\bf p}$ dominates the other terms involving ${\bf p}$ in Eq.~(\ref{eq-p-2}), we can neglect these other terms and this equation can be inverted to obtain
\be
{\bf p} \simeq \frac{1}{6D_{\rm r}} \Big[ -V_{\rm sd}\pmb{\nabla}n_{\rm C} + n_{\rm C}\sum_k(2\lambda_k-\zeta_k)\pmb{\nabla}n_k \Big],
\label{p-grad}
\ee
under which circumstances the field $\bf p$ is driven by the gradients of the colloid and species densities.  Substituting this result into Eq.~(\ref{eq-nC-2}) for the density $n_{\rm C}$ of colloidal particles, we find
\bea
&&\partial_t n_{\rm C} +\pmb{\nabla}\cdot\Big\{n_{\rm C}\sum_k \Big[\xi_k +\frac{V_{\rm sd}}{6D_{\rm r}}(2\lambda_k-\zeta_k)\Big] \pmb{\nabla} n_k\nonumber \\
&&\qquad \qquad \qquad -D_{\rm t}^{\rm (eff)} \pmb{\nabla}n_{\rm C}\Big\}= 0 \, ,
\label{eq-nC-3}
\eea
with the effective diffusion coefficient
\be
D_{\rm t}^{\rm (eff)} \equiv D_{\rm t} + \frac{V_{\rm sd}^2}{6D_{\rm r}} ,
\ee
expressing the enhancement of diffusivity due to the self-diffusiophoretic activity \cite{K13}.

In the following, we suppose that the diffusion coefficient is the same for both molecular species: $D\equiv D_{\rm A}=D_{\rm B}$.  Consequently, $n_0 = n_{\rm A}+ n_{\rm B}$ remains uniform during the time evolution if initially so.  Therefore, $n_{\rm B}=n_0-n_{\rm A}$ is known and only $n_{\rm A}$ needs to be determined.  Introducing the notations
\be
a\equiv n_{\rm A} \qquad\mbox{and} \qquad c \equiv n_{\rm C} \, ,
\ee
we have the following coupled equations describing the system:
\bea
&&\partial_t a = D \nabla^2 a - W_{\rm tot} \, , \label{eq-a}\\
&& W_{\rm tot} = c\,  (K \, a -K_0) + K_2 \, a -K_{20} \, ,  \label{eq-W}\\
&&\partial_t c = \pmb{\nabla} \cdot \left[ \left( D_{\rm t}+\tau_{\rm r}V_{\rm sd}^2\right) \pmb{\nabla} c - (\xi+\sigma V_{\rm sd}) c \; \pmb{\nabla} a \right] ,  \label{eq-c}\\
&& V_{\rm sd} = \zeta  a -V_0 \, ,  \label{eq-V}
\eea
with
\bea
&&K \equiv k_++k_- \, , \quad K_0 \equiv k_- n_0 \, , \quad K_2 \equiv k_{+2}+k_{-2} \, , \nonumber \\
&&K_{20} \equiv k_{-2} n_0 \, , \quad \tau_{\rm r} \equiv (6 D_{\rm r})^{-1} \, , \quad \xi \equiv \xi_{\rm A}-\xi_{\rm B} \, ,\nonumber \\
&&\lambda \equiv \lambda_{\rm A}-\lambda_{\rm B} \, , \quad \zeta \equiv \zeta_{\rm A}-\zeta_{\rm B}=\varsigma(\kappa_+^{\rm c}+\kappa_-^{\rm c}) \, , \nonumber \\
&& V_0 \equiv - \zeta_{\rm B} n_0 =  \varsigma \, \kappa_-^{\rm c} \, n_0 \, , \quad \sigma \equiv \tau_{\rm r} (2\lambda-\zeta) \, .
\eea
Moreover, consistency with the existence of equilibrium requires that $\zeta/K=V_0/K_0=\varsigma/\Gamma=\chi$ is equal to the diffusiophoretic parameter~(\ref{chi}) that is the ratio self-diffusiophoretic velocity~(\ref{Vsd}) and the leading term of the overall reaction rate~(\ref{W_rxn}).

For this system, there exists a uniform nonequilibrium steady state, where $c$ keeps its initial uniform value $c_0$ and the molecular density is also uniform at the value
\be
a_0 = \frac{c_0\, K_0 + K_{20}}{c_0\, K + K_2} ,
\ee
in order to satisfy the stationary condition $W_{\rm tot}=0$.  For this molecular concentration $a=a_0$, we notice that the rate $K a - K_0$ of the catalytic reaction on the colloids is not vanishing under the nonequilibrium condition $k_+/k_- \neq k_{+2}/k_{-2}$.

\begin{figure*}[ht]
\centerline{\scalebox{0.5}{\includegraphics{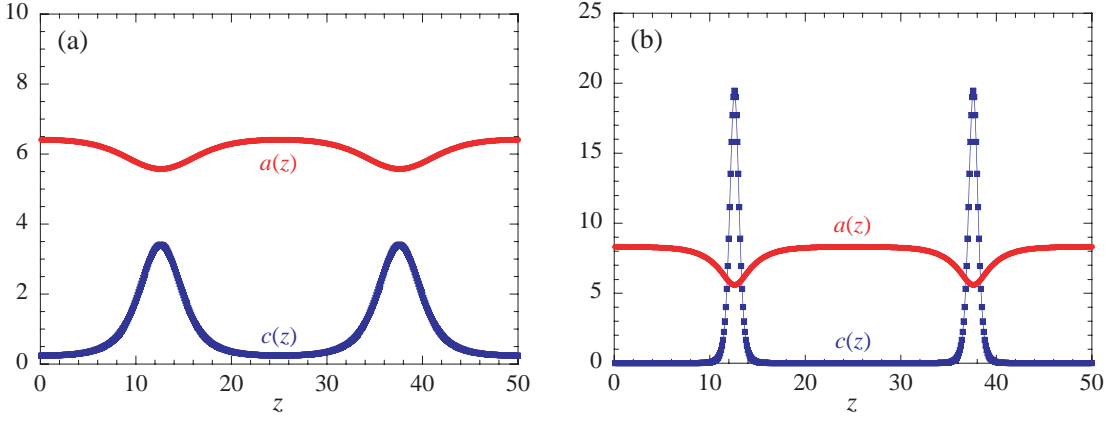}}}
\caption{Nonequilibrium steady state of the one-dimensional system for the parameter values~(\ref{param}): (a)~with $K_{20}=2$; (b)~with $K_{20}=2.5$.}
\label{fig2}
\end{figure*}

\begin{figure*}[ht]
\centerline{\scalebox{0.75}{\includegraphics{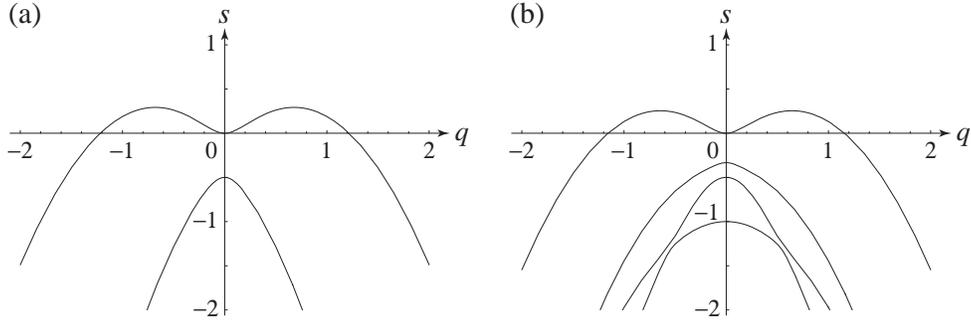}}}
\caption{Dispersion relations of linear stability analysis for $K_{20}=3$ obtained: (a) with the approximation~(\ref{dr2}); (b) by truncating Eq.~(\ref{matrix-f}) into a $5\times 5$ matrix.}
\label{fig3}
\end{figure*}

To analyze the stability of this homogeneous steady state, for simplicity we consider a one-dimensional system where the fields $a$ and $c$ only depend on the variable $z$.  Accordingly, the gradients $\pmb{\nabla}$ can be replaced by partial derivatives $\partial_z$ in Eqs.~(\ref{eq-a}) and~(\ref{eq-c}).  The set of Eqs.~(\ref{eq-a})-(\ref{eq-V}) is then numerically integrated by spatial discretization over the grid $z=i\, \Delta z$ with $i=1,2,...,M$ with $\Delta z=0.1$ and $M=500$.  The integration is performed with a Runge-Kutta algorithm of varying order 4-5 over a long enough time interval to reach a steady state.  Figure~\ref{fig2} shows numerical results for the parameter values
\bea
&& c_0 = 1 \, , \quad n_0= 10 \, , \quad D=1 \, , \quad D_{\rm t}=1 \, , \quad \tau_{\rm r}=1 \, ,\nonumber \\
&& \xi=-3 \, , \quad \sigma=-2 \, , \quad V_0=0.5 \, , \quad \zeta=0.1\, , \nonumber\\
&&  K=0.2 \, , \quad K_0=1 \, , \quad K_2=0.3 \, ,
\label{param}
\eea
and increasing values of $K_{20}$.  We observe the formation of clusters of colloidal motors in regions where the fuel A is depleted, as expected.

The threshold of this clustering instability can be found from a linear stability analysis.  Linearizing the equations around the uniform steady state, we find that the perturbations obey
\be
\partial_t
\left(
\begin{array}{c}
\delta a \\
\delta c \\
\end{array}
\right)
=
\left(
\begin{array}{cc}
D\partial_z^2-\tilde K&- w \\
 -\rho\,\partial_z^2 &D_{\rm t}^{\rm (eff)}\partial_z^2\\
\end{array}
\right)
\left(
\begin{array}{c}
\delta a \\
\delta c \\
\end{array}
\right),
\label{M_eq_lin_stab}
\ee
with
\bea
&&\tilde K \equiv c_0 \, K + K_2 \, , \quad w\equiv K\, a_0 - K_0 \, , \nonumber\\
&& \rho \equiv c_0 \, (\xi+ \sigma V_{\rm sd}) \, , \quad D_{\rm t}^{\rm (eff)} \equiv D_{\rm t}+ \tau_{\rm r} V_{\rm sd}^2 \, , \nonumber \\
&& V_{\rm sd} \equiv \zeta \, a_0 - V_0 \, .
\eea
Supposing that the perturbations behave as $\delta a, \delta c \sim \exp(iqz+st)$, we obtain the dispersion relations
\bea
s_{\pm}(q) &=& -\frac{1}{2}\Big[ \tilde K + (D+D_{\rm t}^{\rm (eff)})q^2\Big]  \nonumber\\
&&\pm \frac{1}{2}\sqrt{\Big[ \tilde K + (D-D_{\rm t}^{\rm (eff)})q^2\Big]^2-4\, \rho\, w \,  q^2} \, .
\label{dr2}
\eea
For the parameter values~(\ref{param}), we have that $\rho <0$ and $w>0$, so that both dispersion relations are real.  They are depicted in Fig.~\ref{fig3}(a) for $K_{20}=3$ beyond the threshold.  The leading dispersion relation is associated with the conserved unstable mode of the colloidal motors because $s_+(0)=0$.  The subleading dispersion relation is associated with the reactive mode of molecular species because $s_-(0)=-\tilde K$.  We notice that, since $\tilde K >0$, there is no possibility for a Hopf bifurcation to oscillatory behavior, which would be the case if the eigenvalues $s_{\pm}$ were complex.  We also note that there is no wavelength selection at the level of linear stability analysis in this clustering instability.

Therefore, instability manifests itself if
\be
\rho \, w + \tilde K \, D_{\rm t}^{\rm (eff)} < 0 \, ,
\ee
and the threshold is given by the condition
\be
\rho \, w + \tilde K \, D_{\rm t}^{\rm (eff)} = 0 \, ,
\ee
which leads to the value $K_{20}\simeq 1.89817$ for the parameter values~(\ref{param}).

The dispersion relations can also be obtained from the evolution equation~(\ref{master}) for the distribution function. Supposing that $f=f(z,\theta)$ and $a=a(z)$, we have the equation
\bea
\partial_t f +\partial_z \left(V_{{\rm d}z} \, f-D_{\rm t}\partial_zf\right) &=& \frac{D_{\rm r}}{\sin\theta}\, \partial_{\theta} \left( \sin\theta \, \partial_{\theta} f \right) \nonumber \\
&&+ 2 \lambda \, \cos\theta \, f \, \partial_z a  \, ,
\label{master-z}
\eea
where
\be
V_{{\rm d}z} = V_{\rm sd}\, \cos\theta +\left[ \xi + \varepsilon \left(\cos^2\theta - \frac{1}{3}\right)\right] \partial_z a \, ,
\ee
which is coupled to the equation for the concentration field $a$ with $c=\int f \, d^2u$.

The linear stability analysis can be carried out for Eq.~(\ref{master-z}) coupled to Eq.~(\ref{eq-a}) with the rate~(\ref{eq-W}) in a similar manner to that for Eqs.~(\ref{eq-a})-(\ref{eq-V}).  This analysis is presented in App.~\ref{app:LinStab-f}.  The dispersion relations can be computed by truncating the infinite matrix~(\ref{matrix-f}) in order to obtain the eigenvalues as a function of the wave number $q$. The result converges to the dispersion relations shown in Fig.~\ref{fig3}(b).  The convergence occurs faster for the leading dispersion relation than for the subleading ones.  For the chosen parameter values, we can see that the approximation where we suppose that the vector field $\bf p$ is driven by the gradients (which corresponds to truncating to a $2\times 2$ matrix) constitutes a good approximation to describe the instability.  Indeed, the leading dispersion relation of Fig.~\ref{fig3}(a) is already very close to that in Fig.~\ref{fig3}(b).

The conclusion is that Eqs.~(\ref{eq-a})-(\ref{eq-V}) provide a robust description of the clustering instability and of the emerging patterns.

\section{Conclusion}\label{sec:Conclusion}

Autonomous motion is not possible at equilibrium and active matter relies on the presence of nonequilibrium constraints to drive the system out of equilibrium. As a result the theoretical formulations provided by nonequilibrium thermodynamics and statistical mechanics are a natural starting point for the description of such systems.

Many of the active matter systems currently under study involve active agents such as molecular machines or self-propelled colloidal particles with linear dimensions ranging from tens of nanometers to micrometers. The transition from microscopic to macroscopic description for fluids containing active agents of such sizes takes place in the upper range of this scale. Suspensions of active colloidal particles are interesting in this connection since, as described earlier in this paper, the colloidal particles are large compared to the molecules of the medium in which they reside. The dynamics of the suspension can then be described by considering the equations for the positions, velocities and orientations of the colloidal particles in the medium, or through field equations that describe the densities of these particles.

Nonequilibrium thermodynamics provides a set of principles that these systems must obey. Most important among these is microreversibility that stems from the basic time reversal character of the microscopic dynamics. On the macroscale this principle manifests itself in Onsager's reciprocal relations that govern what dynamical processes are coupled and how they are described. For example, for single Janus particles propelled by a self-diffusiophoretic mechanism, microreversibility implies the existence of reciprocal effect where the reaction rate depends on an applied external force~\cite{GK17,GK18a,GK19,HSGK18}.

This paper extended the nonequilibrium thermodynamics formulation to the collective dynamics of ensembles of diffusiophoretic Janus colloids. In particular we considered Janus colloids driven by both self-diffusiophoresis arising from reactions on the motor catalytic surface as well as motion arising from an external concentration gradient. This latter contribution is essential for the extension of the theory to collective motor dynamics. The resulting formulation is consistent with microreversibility and an expression for the entropy production is provided. From this general formulation of collective dynamics one can show that if an external force and torque are applied to the system the overall reaction rate depends on the applied force. In addition a stability analysis of the equations governing the collective behavior predicts the existence of a clustering instability seen in many experiments of Janus colloids.  Such considerations can be extended to ensembles of thermophoretic Janus colloids \cite{GK19JSM}.

%%%%%%%%%%%%%%%%%%%%%%%%%%%%%%%%%%%%%%%%%%%\\
\vskip 0.5 cm

\appendix

\section{Force and torque on a colloidal motor}\label{app:FT_motor}

We suppose that molecular diffusivity is large enough so that the concentration fields adopt a quasi-stationary profile in the vicinity of every Janus particle.  Accordingly, the concentrations fields should obey the following equations:
\bea
&&\nabla^2 c_k = 0 \, , \\
&& D_k \, (\partial_r c_k)_R = -\nu_k \, H^{\rm c}(\theta) \, (\kappa_+^{\rm c}\, c_{\rm A}-\kappa_-^{\rm c}\, c_{\rm B})_R \, , \\
&& (\pmb{\nabla}c_k)_{\infty} = {\bf g}_k \, ,
\eea
for the two reacting species $k={\rm A},{\rm B}$, where ${\bf g}_k$ is the concentration gradient of species $k$ at large distances from the center of the catalytic particle.  In Ref.~\cite{GK18a}, we considered the special case where ${\bf g}_k=0$, so that the concentration fields are uniform at large distances.  The upper hemisphere is catalytic, while the lower one is noncatalytic.  The axis of the Janus particle is oriented from the noncatalytic towards the catalytic hemisphere and taken along the $z$-axis in the frame of the particle.

We introduce the fields
\bea
&&\Phi \equiv \ell (D_{\rm A}\, c_{\rm A} + D_{\rm B}\, c_{\rm B}) \, , \label{Phi}\\
&&\Psi \equiv \ell^2 (\kappa_+^{\rm c}\, c_{\rm A} -\kappa_-^{\rm c}\, c_{\rm B}) \, , \label{Psi}
\eea
in terms of the characteristic length of the reaction
\be
\ell \equiv \left(\frac{\kappa_+^{\rm c}}{D_{\rm A}} + \frac{\kappa_-^{\rm c}}{D_{\rm B}}\right)^{-1} .
\label{ell}
\ee
The fields $\Phi$ and $\Psi$ have the units of sec$^{-1}$ and the concentration fields are recovered from them by
\bea
&& c_{\rm A} = \frac{1}{D_{\rm A}} \left( \frac{\kappa_-^{\rm c}}{D_{\rm B}} \, \Phi + \frac{1}{\ell}\, \Psi\right) , \\
&& c_{\rm B} = \frac{1}{D_{\rm B}} \left( \frac{\kappa_+^{\rm c}}{D_{\rm A}} \, \Phi - \frac{1}{\ell}\, \Psi\right) .
\eea
 Similar expressions hold for the concentration gradients at large distances: ${\bf g}_{\rm A}$, ${\bf g}_{\rm B}$, ${\bf g}_{\Phi}$, and ${\bf g}_{\Psi}$.

The fields~(\ref{Phi}) and~(\ref{Psi}) obey the equations
\bea
&&\nabla^2 \Phi = 0 \, , \\
&&(\partial_r \Phi)_R = 0 \, , \\
&& (\pmb{\nabla}\Phi)_{\infty} = {\bf g}_{\Phi} \, ,
\eea
and
\bea
&&\nabla^2 \Psi = 0 \, , \\
&&(\partial_r \Psi)_R = \ell^{-1} H^{\rm c}(\theta)\, (\Psi)_R \, , \\
&& (\pmb{\nabla}\Psi)_{\infty} = {\bf g}_{\Psi} \, ,
\eea
where $H^{\rm c}(\theta)$ is the Heaviside function of the catalytic hemisphere.

The solution of the equations for $\Phi$ is given by
\bea
\Phi=\Phi_g &=& \Phi_0 + {\bf g}_{\Phi}\cdot{\bf r} \left[ 1 + \frac{1}{2}\left(\frac{R}{r}\right)^3\right] ,\\
\Phi_0 &=& \ell(D_{\rm A}\, \bar{c}_{\rm A} +D_{\rm B}\, \bar{c}_{\rm B})\, ,
\label{Phi-g}
\eea
which obeys reflective boundary conditions on the sphere $r=R$ and represents a gradient at large distances. Defining ${\rm Da} \equiv R/\ell$ to be the dimensionless Damk\"ohler number of the reaction on the spherical catalytic particle, the field $\Psi$ can be decomposed as
\be
\Psi=\Psi_g -{\rm Da} \, \Psi_0 \, {\cal F}
\ee
in terms of a field similar to Eq.~(\ref{Phi-g})
\bea
\Psi_g &=& \Psi_0 + {\bf g}_{\Psi}\cdot{\bf r} \left[ 1 + \frac{1}{2}\left(\frac{R}{r}\right)^3\right],
\label{Psi-g}\\
\Psi_0 &\equiv& \ell^2 (\kappa_+^{\rm c}\, \bar{c}_{\rm A} -\kappa_-^{\rm c}\, \bar{c}_{\rm B}) \, , \label{Psi0}
\eea
and another field ${\cal F}$ obeying
\bea
&&\nabla^2 {\cal F} = 0 \, , \\
&&R\, (\partial_r {\cal F})_R = H^{\rm c}(\theta) \left[ {\rm Da}\, ({\cal F})_R -\frac{(\Psi_g)_R}{\Psi_0}\right]  , \label{rbc}\\
&& (\pmb{\nabla}{\cal F})_{\infty} = 0 \, .
\eea
In the equations above the concentrations $\bar{c}_k$ may be considered as their extrapolations to the center of the Janus particle or the mean concentrations at that position in a dilute suspension of Janus particles. Similarly, ${\bf g}_k$ may also be considered as the mean gradients of concentrations at the location of the Janus particle in a dilute suspension.

The diffusiophoretic force and torque are thus given by the following expressions
\bea
&&{\bf F}_{\rm d} = \frac{6\pi\eta R}{1+3b/R} \, \Big[\left[\frac{3}{2} \, \overline{b_{\rm A} {\boldsymbol{\mathsf 1}}_{\bot}}^{\rm s}\cdot{\bf g}_{\rm A} + \frac{3}{2} \, \overline{b_{\rm B}  {\boldsymbol{\mathsf 1}}_{\bot}}^{\rm s}\cdot{\bf g}_{\rm B} \right.\nonumber \\
&&\left.+R(\kappa_+^{\rm c}\, \bar{c}_{\rm A} -\kappa_-^{\rm c}\, \bar{c}_{\rm B})\overline{\left(\frac{b_{\rm B}}{D_{\rm B}}-\frac{b_{\rm A}}{D_{\rm A}}\right)  {\boldsymbol{\mathsf 1}}_{\bot}\cdot\pmb{\nabla}{\cal F}}^{\rm s}\right]
\label{Fd4}
\eea
and
\bea
&&{\bf T}_{\rm d}= \frac{12\pi\eta  R}{1+3b/R} \, \left[\frac{3R}{2} \, \overline{b_{\rm A}  {\bf n}}^{\rm s}\times{\bf g}_{\rm A} + \frac{3R}{2} \, \overline{b_{\rm B} {\bf n}}^{\rm s}\times{\bf g}_{\rm B}\right.\nonumber \\
&&\left.+R(\kappa_+^{\rm c}\, \bar{c}_{\rm A} -\kappa_-^{\rm c}\, \bar{c}_{\rm B}) \overline{\left(\frac{b_{\rm B}}{D_{\rm B}}-\frac{b_{\rm A}}{D_{\rm A}}\right)  {\bf r}\times\pmb{\nabla}{\cal F}}^{\rm s} \right] .
\label{Td4}
\eea

Next, the field $\cal F$ can be expanded in Legendre polynomials.  Since it obeys Laplace's equation and is vanishing at large distances, we find that
\bea
&&{\cal F}(r,\theta,\varphi) = \sum_{l=0}^{\infty} a_l \, P_l(\cos\theta) \, \left(\frac{R}{r}\right)^{l+1} \nonumber\\
&&\quad +  \frac{3R}{2\Psi_0} \, g_{\Psi z} \sum_{l=0}^{\infty} b_l \, P_l(\cos\theta) \, \left(\frac{R}{r}\right)^{l+1} \nonumber\\
&& \quad+  \frac{3R}{2\Psi_0} \, \sqrt{2}\, \left(g_{\Psi x}\, \cos\varphi + g_{\Psi y}\, \sin\varphi \right)\times \nonumber \\
 &&\quad\qquad \qquad \times \sum_{l=1}^{\infty} c_l \, \frac{P_l^1(\cos\theta)}{\sqrt{l(l+1)}} \, \left(\frac{R}{r}\right)^{l+1}
\label{F_expansion}
\eea
with
\bea
&& a_l \equiv \left({\boldsymbol{\mathsf M}}^{-1}\cdot\pmb{\cal A}\right)_l \, , \label{al}\\
&& b_l \equiv \left({\boldsymbol{\mathsf M}}^{-1}\cdot\pmb{\cal B}\right)_l \, , \label{bl}\\
&& c_l \equiv  \left({\boldsymbol{\mathsf N}}^{-1}\cdot\pmb{\cal C}\right)_l \, , \label{cl}\\
&& {\cal A}_l \equiv \int _0^1 d\xi \, P_l(\xi) \, , \\
&& {\cal B}_l \equiv \int _0^1 d\xi \, P_1(\xi) \, P_l(\xi) \, , \\
&& {\cal C}_l \equiv \int _0^1 d\xi \, \frac{P_1^1(\xi)}{\sqrt{2}} \, \frac{P_l^1(\xi)}{\sqrt{l(l+1)}} \, , \\
&& M_{ll'} \equiv 2 \, \frac{l+1}{2l+1} \, \delta_{ll'} + {\rm Da} \, \int_0^1 d\xi \, P_l(\xi)\, P_{l'}(\xi) \, , \\
&& N_{ll'} \equiv 2 \, \frac{l+1}{2l+1} \, \delta_{ll'} \nonumber \\
&& \qquad + {\rm Da} \, \int_0^1 d\xi \, \frac{P_l^1(\xi)}{\sqrt{l(l+1)}} \, \frac{P_{l'}^1(\xi)}{\sqrt{l'(l'+1)}} \, .
\eea

First, we calculate the force~(\ref{Fd4}) in order to obtain the corresponding velocity ${\bf V}_{\rm d}$.  We have that
\bea
\overline{b_k  {\bf n} {\bf n}}^{\rm s} &=& \frac{1}{6}\, (b_k^{\rm c}+b_k^{\rm n}) \, {\boldsymbol{\mathsf 1}} \, ,\\
\overline{b_k  {\boldsymbol{\mathsf 1}}_{\bot}}^{\rm s} &=& \frac{1}{3}\, (b_k^{\rm c}+b_k^{\rm n}) \, {\boldsymbol{\mathsf 1}} \, ,
\eea
and, furthermore,
\be
{\bf 1}_x\cdot\overline{b_k{\boldsymbol{\mathsf 1}}_{\bot}\cdot\pmb{\nabla}{\cal F}}^{\rm s} = -\frac{1}{2}\left(\gamma^{\rm c}b_k^{\rm c}+\gamma^{\rm n}b_k^{\rm n}\right)\frac{g_{\Psi x}}{\Psi_0} \, , 
\ee
\bea
{\bf 1}_y\cdot\overline{b_k{\boldsymbol{\mathsf 1}}_{\bot}\cdot\pmb{\nabla}{\cal F}}^{\rm s} &=& -\frac{1}{2}\left(\gamma^{\rm c}b_k^{\rm c}+\gamma^{\rm n}b_k^{\rm n}\right)\frac{g_{\Psi y}}{\Psi_0}\, , \\
{\bf 1}_z\cdot\overline{b_k{\boldsymbol{\mathsf 1}}_{\bot}\cdot\pmb{\nabla}{\cal F}}^{\rm s} &=& -\frac{1}{2R} \left(\alpha^{\rm c}b_k^{\rm c}+\alpha^{\rm n}b_k^{\rm n}\right) \nonumber \\ &&-\frac{1}{2}\left(\beta^{\rm c}b_k^{\rm c}+\beta^{\rm n}b_k^{\rm n}\right)\frac{g_{\Psi z}}{\Psi_0}\, ,
\eea
in terms of the integrals
\be
\alpha^{h}\equiv \int d\theta \, H^h(\theta)\, \sin^2\theta \, \frac{dA}{d\theta}\, , 
\ee
\bea
&& \beta^{h}\equiv \frac{3}{2}\, \int d\theta \, H^h(\theta)\, \sin^2\theta \, \frac{dB}{d\theta}\, , \\
&& \gamma^{h}\equiv -\frac{3\sqrt{2}}{4} \int d\theta  H^h(\theta) \Big(\sin\theta  \cos\theta  \frac{dC}{d\theta} + C\Big) , \quad
\eea
where
\bea
&& A \equiv \sum_{l=0}^{\infty} a_l \, P_l(\cos\theta) \, , \\
&& B \equiv \sum_{l=0}^{\infty} b_l \, P_l(\cos\theta) \, , \\
&& C \equiv \sum_{l=1}^{\infty} c_l \, \frac{P_l^1(\cos\theta)}{\sqrt{l(l+1)}} \, .
\eea

Then we calculate the torque~(\ref{Td4}) to get the corresponding angular velocity $\pmb{\Omega}_{\rm d}$.  We have that
\be
\overline{b_k {\bf n}}^{\rm s} = \frac{1}{4}\, (b_k^{\rm c}-b_k^{\rm n}) \, {\bf u} \, ,
\ee
and
\be
{\bf 1}_x\cdot\overline{b_k{\bf r}\times\pmb{\nabla}{\cal F}}^{\rm s} =\frac{R}{2}\left(\delta^{\rm c}b_k^{\rm c}+\delta^{\rm n}b_k^{\rm n}\right)\frac{g_{\Psi y}}{\Psi_0} \, , 
\ee
\be
{\bf 1}_y\cdot\overline{b_k{\bf r}\times\pmb{\nabla}{\cal F}}^{\rm s} = -\frac{R}{2}\left(\delta^{\rm c}b_k^{\rm c}+\delta^{\rm n}b_k^{\rm n}\right)\frac{g_{\Psi x}}{\Psi_0}\, , 
\ee
\be
{\bf 1}_z\cdot\overline{b_k{\bf r}\times\pmb{\nabla}{\cal F}}^{\rm s} = 0 \, ,
\ee
with
\be
\delta^{h}\equiv -\frac{3\sqrt{2}}{4}\, \int d\theta \, H^h(\theta)\, \left(\sin\theta \, \frac{dC}{d\theta} + \cos\theta \, C\right)\, ,
\ee
so that
\be
\overline{b_k{\bf r}\times\pmb{\nabla}{\cal F}}^{\rm s} = \frac{R}{2\Psi_0}\left(\delta^{\rm c}b_k^{\rm c}+\delta^{\rm n}b_k^{\rm n}\right) {\bf u}\times {\bf g}_{\Psi} \, .
\ee

With the following coefficients associated with diffusiophoresis,
\be
Y^h \equiv \frac{b_{\rm A}^{h}}{D_{\rm A}}-\frac{b_{\rm B}^{h}}{D_{\rm B}} \qquad\mbox{for}\qquad h={\rm c,n} \, ,
\ee
the diffusiophoretic linear and angular velocities can be expressed as
\begin{widetext}
\bea
{\bf V}_{\rm d} = \frac{{\bf F}_{\rm d}}{\gamma_{\rm t}} &=& \frac{1}{1+2b/R}\Bigl\{
 \frac{1}{2} \left( \alpha^{\rm c} Y^{\rm c} + \alpha^{\rm n} Y^{\rm n}\right) (\kappa_+^{\rm c}\, \bar{c}_{\rm A}-\kappa_-^{\rm c}\, \bar{c}_{\rm B})\, {\bf u}+\frac{1}{2}(b_{\rm A}^{\rm c}+b_{\rm A}^{\rm n}) \, {\bf g}_{\rm A} + \frac{1}{2}(b_{\rm B}^{\rm c}+b_{\rm B}^{\rm n}) \, {\bf g}_{\rm B}  \nonumber\\
&+& \frac{R}{2}\left( \gamma^{\rm c} Y^{\rm c} +  \gamma^{\rm n} Y^{\rm n} \right) (\kappa_+^{\rm c}\, {\bf g}_{\rm A} -\kappa_-^{\rm c}\, {\bf g}_{\rm B})+ \frac{R}{2}\left[ (\beta^{\rm c}-\gamma^{\rm c})Y^{\rm c}  + (\beta^{\rm n}-\gamma^{\rm n})Y^{\rm n}\right] (\kappa_+^{\rm c}\, {\bf g}_{\rm A} -\kappa_-^{\rm c} {\bf g}_{\rm B})\cdot {\bf u}\, {\bf u}\Bigr\} , \quad \label{eq:velocity}\\
\pmb{\Omega}_{\rm d} = \frac{{\bf T}_{\rm d}}{\gamma_{\rm r}}&=& \frac{9}{16R} \left[\left(b_{\rm A}^{\rm c}-b_{\rm A}^{\rm n}\right) {\bf u}\times {\bf g}_{\rm A} + \left(b_{\rm B}^{\rm c}-b_{\rm B}^{\rm n}\right) {\bf u}\times{\bf g}_{\rm B} \right]+ \frac{3}{4} \left( \delta^{\rm c} Y^{\rm c} +  \delta^{\rm n}Y^{\rm n} \right) {\bf u}\times (\kappa_+^{\rm c}\, {\bf g}_{\rm A} -\kappa_-^{\rm c}\, {\bf g}_{\rm B}) \, .
\label{eq:omega}
\eea
\end{widetext}
By defining the parameters,
\begin{eqnarray}
\zeta_{\rm A}&=&\varsigma \kappa_+^{\rm c} ,\quad \zeta_{\rm B}=-\varsigma \kappa_-^{\rm c} ,\nonumber \\
&&\varsigma=\frac{1}{2}\frac{\sum_{h}\alpha^h Y^h}{1+2b/R}, \label{eq:v-Omega-parameters1}\\
\xi_{\rm A}&=&\frac{1}{2}\frac{\sum_h b_{\rm A}^h}{1+2b/R} +\Xi  \kappa_+^{\rm c}, \quad \xi_{\rm B}=\frac{1}{2}\frac{\sum_h b_{\rm B}^h}{1+2b/R}-\Xi  \kappa_-^{\rm c}, \nonumber \\
&& \Xi= \frac{R}{6} \frac{\sum_h (\beta^h + 2\gamma^h)Y^h}{1+2b/R},
\label{eq:v-Omega-parameters2}\\
\varepsilon_{\rm A}&=& E \kappa_+^{\rm c}, \quad \varepsilon_{\rm B}= -E \kappa_-^{\rm c}, \nonumber \\
&&E=\frac{R}{2} \frac{\sum_h(\beta^h-\gamma^h)Y^h}{1+2b/R},  \label{eq:v-Omega-parameters3}\\
\lambda_{\rm A}&=& \frac{9(b_{\rm A}^{\rm c}-b_{\rm A}^{\rm n})}{16R} +\Lambda \kappa_+^{\rm c}, \quad \lambda_{\rm B}= \frac{9(b_{\rm B}^{\rm c}-b_{\rm B}^{\rm n})}{16R} -\Lambda \kappa_-^{\rm c},\nonumber \\
&&\Lambda= \frac{3}{4} \sum_h \delta^h Y^h,\label{eq:v-Omega-parameters4}
\end{eqnarray}
the linear and angular velocities in Eqs.~(\ref{eq:velocity}) and~(\ref{eq:omega}) can be written in the forms given in Eqs.~(\ref{Vd}) and (\ref{Omegad}) in the main text.

Moreover, we can also calculate the overall reaction rate~(\ref{W_rxn}).
According to the boundary condition
\be
D_{\rm A} \, (\partial_r c_{\rm A})_R = H^{\rm c}(\theta)\, (\kappa_+^{\rm c}\, c_{\rm A}-\kappa_-^{\rm c}\, c_{\rm B})_R \, ,
\ee
the reaction rate on the Janus particle is equivalently given by
\be
W_{\rm rxn}=\int_{r=R} D_{\rm A} \, (\partial_r c_{\rm A})_R\, dS ,
\ee
with $dS = R^2 \, d\cos\theta\, d\varphi$. The concentration field $c_{\rm A}$ is again decomposed in terms of $\Phi$ and $\Psi$.
The contributions from the terms $\Phi=\Phi_g$ and $\Psi_g$ are vanishing, so that there remains
\be
W=-\frac{\rm Da}{\ell}\, \Psi_0 \int_{r=R} (\partial_r {\cal F})_R\, dS \, .
\ee
Using the expansion~(\ref{F_expansion}), we obtain the overall reaction rate~(\ref{W_rxn}) with the rate constants
\be\label{k_rxn}
k_{\pm} = 4\pi R^2 a_0 \kappa_{\pm}
\ee
and the parameter
\be\label{varpi}
\varpi = \frac{3Rb_0}{2a_0} \, ,
\ee
where the coefficients $a_0$ and $b_0$ are given by Eqs.~(\ref{al}) and~(\ref{bl}) with $l=0$.

%%%%%%%%%%%%%%%%%%%%%%%%%%%%%%%%%%%%%%%%%%%%%%%%%

\section{Determination of Onsager's linear response coefficients}\label{app:Onsager}

In Euclidean space, the contravariant $a^i$ and covariant $a_i$ components of a vector ${\bf a}\in{\mathbb R}^3$ coincide: $a_i=a^i$.  However, in spherical coordinates, the contravariant $a_{\rm r}^i$ and covariant $a_{{\rm r}i}$ components of a rotational vector ${\bf a}_{\rm r}\in{\mathbb R}^2$ are related to each other according to $a_{{\rm r}i}=\sum_{j=\theta,\varphi} g_{ij} a_{\rm r}^j$ in terms of the metric~(\ref{metric}).  Therefore, the scalar product between two rotational vectors ${\bf a}_{\rm r},{\bf b}_{\rm r}\in{\mathbb R}^2$ has the following equivalent forms, ${\bf a}_{\rm r}\bullet{\bf b}_{\rm r}= \sum_{i=\theta,\varphi} a_{{\rm r}i} b_{\rm r}^i = \sum_{i=\theta,\varphi} a_{\rm r}^i b_{{\rm r}i}$.   The inverse of the metric~(\ref{metric}) associated with the spherical coordinates is given by
\be
(g^{ij}) = (g_{ij})^{-1} =  \left(
\begin{array}{cc}
1 & 0 \\
0 & \frac{1}{\sin^2\theta}
\end{array}
\right)
\ee
and its determinant by
\be
g={\rm det} (g_{ij}) = \sin^2\theta \, .
\ee
Hence, the element of angular integration can be written as $d^2u=\sqrt{g}\, d^2q=\sin\theta\, d\theta\, d\varphi$.  Using contravariant components, the gradient of some function $X$ is given by \cite{B65}
\be\label{grad_r-i}
({\rm grad}_{\rm r}X)^i=\sum_j g^{ij} \, \frac{\partial X}{\partial q^j}
\ee
and the divergence of some vector $\pmb{X}_{\rm r}$ by
\be\label{div_r-i}
{\rm div}_{\rm r} \,\pmb{X}_{\rm r}= \sum_i \frac{1}{\sqrt{g}} \, \frac{\partial}{\partial q^i}\left( {X}_{\rm r}^{i} \sqrt{g}\right) ,
\ee
which leads to Eqs.~(\ref{grad_r}) and~(\ref{div_r}) with the metric~(\ref{metric}) of spherical coordinates.

Using the chemical potentials (\ref{mu-k})-(\ref{mu-C}) and the assumptions~(\ref{Fick}) and~(\ref{Lrk=0}), Eq.~(\ref{J=LA}) becomes
\be
\left(\begin{array}{c}
w \\
\pmb{\jmath}_{\rm A}\\
\pmb{\jmath}_{\rm B}\\
{\bf J}_{\rm C}
\end{array}
\right)
=
\left(\begin{array}{cccc}
L_{\rm rr} & {\bf 0} & {\bf 0} &  {\bf L}_{\rm rC}  \\
{\bf 0} & {\boldsymbol{\mathsf L}}_{\rm AA} & {\boldsymbol{\mathsf 0}}& {\boldsymbol{\mathsf L}}_{\rm AC} \\
{\bf 0} & {\boldsymbol{\mathsf 0}} & {\boldsymbol{\mathsf L}}_{\rm BB}& {\boldsymbol{\mathsf L}}_{\rm BC} \\
{\bf L}_{\rm Cr} & {\boldsymbol{\mathsf L}}_{\rm CA} & {\boldsymbol{\mathsf L}}_{\rm CB}& {\boldsymbol{\mathsf L}}_{\rm CC}
\end{array}
\right)\cdot
\left(\begin{array}{c}
A_{\rm rxn} \\
-\frac{\pmb{\nabla}n_{\rm A}}{n_{\rm A}}\\
-\frac{\pmb{\nabla}n_{\rm B}}{n_{\rm B}}\\
-{\rm grad}\,\frac{\mu_{\rm C}}{k_{\rm B}T}
\end{array}
\right) ,
\label{lin2}
\ee
which implies that
\be
w = L_{\rm rr} \, A_{\rm rxn} - \sum_{\bf u} {\bf L}_{\rm rC}\cdot{\rm grad}\,\frac{\mu_{\rm C}}{k_{\rm B}T} \, , 
\ee
\bea
&& \pmb{\jmath}_{\rm A} =  - {\boldsymbol{\mathsf L}}_{\rm AA}\cdot\frac{\pmb{\nabla}n_{\rm A}}{n_{\rm A}}  - \sum_{\bf u}{\boldsymbol{\mathsf L}}_{\rm AC}\cdot{\rm grad}\,\frac{\mu_{\rm C}}{k_{\rm B}T} \, , \label{JA-Onsager}\\
&&\pmb{\jmath}_{\rm B} =  - {\boldsymbol{\mathsf L}}_{\rm BB}\cdot\frac{\pmb{\nabla}n_{\rm B}}{n_{\rm B}}  - \sum_{\bf u}{\boldsymbol{\mathsf L}}_{\rm BC}\cdot{\rm grad}\,\frac{\mu_{\rm C}}{k_{\rm B}T} \, ,
\label{JB-Onsager}\\
&&{\bf J}_{\rm C} = {\bf L}_{\rm Cr} \, A_{\rm rxn} - {\boldsymbol{\mathsf L}}_{\rm CA}\cdot\frac{\pmb{\nabla}n_{\rm A}}{n_{\rm A}} - {\boldsymbol{\mathsf L}}_{\rm CB}\cdot\frac{\pmb{\nabla}n_{\rm B}}{n_{\rm B}} \nonumber \\
&&\ \qquad - \sum_{\bf u}{\boldsymbol{\mathsf L}}_{\rm CC}\cdot{\rm grad}\,\frac{\mu_{\rm C}}{k_{\rm B}T} \, ,
\label{JC-Onsager}
\eea
where the sum extends over the different groups $\Delta^2u$ of colloidal motors and
\be
{\rm grad}\, \frac{\mu_{\rm C}}{k_{\rm B}T}=
\frac{1}{f} \left(\begin{array}{c}
\pmb{\nabla}f +f\beta\pmb{\nabla} U_{\rm t} \\
\partial_{\theta}f-f\beta\mu\, {\bf B}\cdot\partial_{\theta}{\bf u} \\
\frac{1}{\sin^2\theta}\, \partial_{\varphi}f  -f\frac{\beta\mu}{\sin^2\theta}\, {\bf B}\cdot\partial_{\varphi}{\bf u}
\end{array}\right) .
\ee

Using the expression~(\ref{JC2}) of the five-dimensional colloidal current density and comparing with the expression~(\ref{JC-Onsager}), we find that the linear response coefficients are given by
\be
{\bf L}_{\rm Cr} = f  \left(
\begin{array}{c}
\chi \, D_{\rm rxn} \, {\bf u} \\
{\bf 0}
 \end{array}
\right) \Delta^2u \ \delta_{\bf uu'} \, , 
\ee
\be
{\boldsymbol{\mathsf L}}_{\rm CA} =-  f  \, n_{\rm A} \left(
\begin{array}{c}
\xi_{\rm A}\, {\boldsymbol{\mathsf 1}} + \varepsilon_{\rm A} \, {\boldsymbol{\mathsf Q}}_{\bf u} \\
\lambda_{\rm A} \, {\rm grad}_{\rm r} {\bf u}
 \end{array}
\right) \Delta^2u \ \delta_{\bf uu'} \, , 
\ee
\be
{\boldsymbol{\mathsf L}}_{\rm CB} =-  f  \, n_{\rm B} \left(
\begin{array}{c}
\xi_{\rm B}\, {\boldsymbol{\mathsf 1}} + \varepsilon_{\rm B} \, {\boldsymbol{\mathsf Q}}_{\bf u} \\
\lambda_{\rm B} \, {\rm grad}_{\rm r} {\bf u}
 \end{array}
\right) \Delta^2u \ \delta_{\bf uu'} \, ,
\ee
and
\be
 {\boldsymbol{\mathsf L}}_{\rm CC} = f  \left(
\begin{array}{cc}
 D_{\rm t} {\boldsymbol{\mathsf 1}} &  {\boldsymbol{\mathsf 0}}   \\
 {\boldsymbol{\mathsf 0}} &  D_{\rm r}  {\boldsymbol{\mathsf 1}}_{\rm r}
 \end{array}
\right) \Delta^2u \ \delta_{\bf uu'} \, .
\ee
Consequently, we find Eqs.~(\ref{w}) and~(\ref{Jk}).

\vskip 0.3 cm

\noindent{Remark.} Interestingly, the assumption~(\ref{Lrk=0}), according to which the reaction rate does not depend on the gradients of molecular densities, can be relaxed by taking instead
\bea
&&{\bf L}_{\rm rA} = -\varpi\,  k_+ \, n_{\rm A}\,  {\bf p}  \, , \\
&&{\bf L}_{\rm rB} = +\varpi \, k_- \, n_{\rm B}\,  {\bf p}  \, ,
\eea
with the polarizability vector~(\ref{p-vec}).
Cross-diffusion between the molecular species A and B may also be included with the coefficients
\be
{\boldsymbol{\mathsf L}}_{\rm AB}={\boldsymbol{\mathsf L}}_{\rm BA}=C \, n_{\rm A} \, n_{\rm B} \, {\boldsymbol{\mathsf 1}} \, .
\ee
In this general case where the matrix of linear response coefficients in Eq.~(\ref{J=LA}) is complete, the reaction rate and the molecular current densities are instead given by
\begin{widetext}
\begin{eqnarray}
w &=& \left(k_+ n_{\rm A}  -  k_- n_{\rm B}\right) n_{\rm C} + \varpi \left(k_+ \pmb{\nabla}n_{\rm A}  -  k_- \pmb{\nabla}n_{\rm B} \right)\cdot{\bf p} - \chi \, D_{\rm rxn} \int {\bf u}\cdot\left(\pmb{\nabla}f+f\beta\pmb{\nabla}U_{\rm t}\right)  d^2u\, , \label{w2}\\
\pmb{\jmath}_{\rm A} &=& -D_{\rm A}\, \pmb{\nabla} n_{\rm A}  -C \, n_{\rm A}\, \pmb{\nabla} n_{\rm B} -\varpi\, k_+ \, n_{\rm A} \, A_{\rm rxn} \, {\bf p} \nonumber\\
&& +\, n_{\rm A} \int\left[\left(\xi_{\rm A}\, {\boldsymbol{\mathsf 1}} + \varepsilon_{\rm A} \, {\boldsymbol{\mathsf Q}}_{\bf u}\right)\cdot\left(\pmb{\nabla}f+f\beta\pmb{\nabla}U_{\rm t}\right)   + \lambda_{\rm A} \, ({\rm grad}_{\rm r} {\bf u})\bullet\left({\rm grad}_{\rm r}  f - f \beta  \mu \, {\bf B}\cdot {\rm grad}_{\rm r} {\bf u}\right)\right]   d^2u\, ,  \label{JA2}\\
\pmb{\jmath}_{\rm B} &=& -D_{\rm B}\, \pmb{\nabla} n_{\rm B}  -C \, n_{\rm B}\, \pmb{\nabla} n_{\rm A} + \varpi\, k_- \, n_{\rm B} \, A_{\rm rxn} \, {\bf p}\nonumber\\
&&+\, n_{\rm B} \int\left[\left(\xi_{\rm B}\, {\boldsymbol{\mathsf 1}} + \varepsilon_{\rm B} \, {\boldsymbol{\mathsf Q}}_{\bf u}\right) \cdot\left(\pmb{\nabla}f+f\beta\pmb{\nabla}U_{\rm t}\right) + \lambda_{\rm B} \, ({\rm grad}_{\rm r} {\bf u})\bullet\left({\rm grad}_{\rm r}  f - f \beta  \mu \, {\bf B}\cdot {\rm grad}_{\rm r}  {\bf u}\right)\right]    d^2u\, . \label{JB2}
\end{eqnarray}
\end{widetext}
Neglecting the last term in the expression~(\ref{w2}), we recover a form compatible with the reaction rate~(\ref{W_rxn}) obtained in Appendix~\ref{app:FT_motor} by direct calculation using the molecular concentration profiles around a single motor.  The scheme has great generality.

%%%%%%%%%%%%%%%%%%%%%%%%%%%%%%%%%%%%%%%%%%%%%%%%%
\section{Linear stability analysis using the colloidal distribution function}\label{app:LinStab-f}

Linearizing Eq.~(\ref{master-z}) for $f$ around a uniform steady state $f_0=c_0/(4\pi)$, we get
\bea
&&\partial_t \delta f =D_{\rm t}\,\partial_z^2 \delta f + \frac{D_{\rm r}}{\sin\theta}\, \partial_{\theta} \left( \sin\theta \, \partial_{\theta} \delta f \right) - V_{\rm sd} \, \cos\theta \, \partial_z\delta f \nonumber \\
&&-f \left[\xi + \varepsilon \left(\cos^2\theta - \frac{1}{3}\right)\right] \partial_z^2 \delta a   + (2 \lambda-\zeta) \, f \, \cos\theta \, \partial_z \delta a  \, ,\nonumber \\
\label{lin-z}
\eea
where $\delta f = f-f_0$ and $\delta a$ is ruled by the first line of matricial Eq.~(\ref{M_eq_lin_stab}).  These linear equations can be solved by expanding $\delta f$ in series of Legendre polynomials as
\be
\delta f = \frac{1}{4\pi} \sum_{l=0}^{\infty} \delta c_l \, P_l(\cos\theta) \, .
\ee
Supposing that $\delta f,\delta a\sim \exp(iqz)$, we find the following coupled equations
\bea
\frac{d\delta c_l}{dt} &=& - \left[ D_{\rm t} q^2 + l(l+1) \, D_{\rm r}\right] \delta c_l  \\
&& -iq \, V_{\rm sd} \left( \frac{l}{2l+1}\, \delta c_{l-1} + \frac{l+1}{2l+3} \, \delta c_{l+1} \right) \nonumber\\
&& + c_0 \left[ q^2 \left( \xi \, \delta_{l,0} + \frac{2}{3}\, \varepsilon\, \delta_{l,2}\right) +iq \, (2\lambda-\zeta) \, \delta_{l,1}\right] \delta a \nonumber
\eea
for $l=0,1,2,...$, and
\be
\frac{d\delta a}{dt} = - (Dq^2 + \tilde K) \, \delta a - w \, \delta c_0 \, .
\ee
In matrix form, we have
\begin{widetext}
\bea
&&\frac{d}{dt}
\left(
\begin{array}{c}
\delta a \\
\delta c_0 \\
\delta c_1 \\
\delta c_2 \\
\delta c_3 \\
\delta c_4 \\
\delta c_5 \\
\vdots
\end{array}
\right)
=
\left(
\begin{array}{ccccccc}
-Dq^2-\tilde K&- w & 0 & 0 & 0 & 0  & \cdots\\
 c_0\xi q^2 &-D_{\rm t}q^2 & -i \frac{V_{\rm sd}}{3} q & 0 & 0 &  0 & \cdots\\
 i c_0 (2\lambda-\zeta) q & -i V_{\rm sd} q & -D_{\rm t}q^2 -2 D_{\rm r} & -i \frac{2V_{\rm sd}}{5} q  & 0 & 0 & \cdots\\
c_0 \frac{2\varepsilon}{3} q^2 & 0 & -i \frac{2V_{\rm sd}}{3} q & -D_{\rm t}q^2 -6 D_{\rm r} & -i \frac{3V_{\rm sd}}{7} q  &  0 & \cdots\\
0 & 0 & 0 & -i \frac{3V_{\rm sd}}{5}q & -D_{\rm t}q^2 -12 D_{\rm r} & -i \frac{4V_{\rm sd}}{9} q  &  \cdots\\
0 & 0 & 0 & 0 & -i \frac{4V_{\rm sd}}{7}q & -D_{\rm t}q^2 -20 D_{\rm r} &  \cdots\\
\vdots & \vdots & \vdots & \vdots & \vdots & \vdots & \ddots\\
\end{array}
\right)
\left(
\begin{array}{c}
\delta a \\
\delta c_0 \\
\delta c_1 \\
\delta c_2 \\
\delta c_3 \\
\delta c_4 \\
\delta c_5 \\
\vdots
\end{array}
\right), \nonumber\\
&&\label{matrix-f}
\eea
\end{widetext}
which can be solved by truncation to obtain the dispersion relations shown in Fig.~\ref{fig3}(b).  If the wave number is vanishing ($q=0$), the matrix in Eq.~(\ref{matrix-f}) have the following successive eigenvalues: $s_-(0)=-\tilde K$ for reaction, $s_+(0)=0$ for translational diffusion, and $s_l(0)=-l(l+1) D_{\rm r}$ with $l=1,2,3,\dots$ for rotational diffusion.  In Fig.~\ref{fig3}(b), they appear in the order $s_+(0)=0 > s_1(0)=-2D_{\rm r} > s_-(0)=-\tilde K > s_2(0)=-6 D_{\rm r} > \cdots$ for the parameter values~(\ref{param}).

%%%%%%%%%%%%%%%%%%%%%%%%%%%%%%%%%%%%%%%%%%%%%%%%%%%
\section*{Acknowledgments}
Research was supported in part by the Natural Sciences and Engineering Research Council of Canada and Compute Canada. Financial support from the Universit\'e libre de Bruxelles (ULB), the Fonds de la Recherche Scientifique~-~FNRS under the Grant PDR~T.0094.16 for the project ``SYMSTATPHYS" is also acknowledged.

%%%%%%%%%%%%%%%%%%%%%%%%%%%%%%%%%%%%%%%%%%%%%%%%%%%

\end{document}